\documentclass[prd,showpacs,aps,amssymb]{revtex4}
\usepackage{graphicx,epsfig} 
\usepackage{bm}

\newcommand{\beq}{\begin{equation}}
\newcommand{\eeq}{\end{equation}}
\newcommand{\ba}{\begin{array}}
\newcommand{\ea}{\end{array}}
\newcommand{\lsim}   {\mathrel{\mathop{\kern 0pt \rlap
  {\raise.2ex\hbox{$<$}}}
  \lower.9ex\hbox{\kern-.190em $\sim$}}}
\newcommand{\gsim}   {\mathrel{\mathop{\kern 0pt \rlap
  {\raise.2ex\hbox{$>$}}}
\lower.9ex\hbox{\kern-.190em $\sim$}}}
\newcommand{\etal}{\emph{et al.}}

\begin{document}

\title{A dip in the UHECR spectrum and the transition from galactic
to extragalactic cosmic rays}

\author{Roberto Aloisio and Veniamin Berezinsky}
 \affiliation{INFN, Laboratori Nazionali del Gran Sasso, I-67010
  Assergi (AQ), Italy}

\author{Pasquale Blasi}
 \affiliation{INAF, Osservatorio Astrofisico di Arcetri, Largo E. Fermi,
  5 - 50125 Firenze, Italy}

\author{Askhat Gazizov}
\affiliation{B.I. Stepanov Institute of Physics of the National
Academy of Sciences of Belarus,\\
F. Skariny Ave.\ 68, 220062 Minsk, Belarus}

\author{Svetlana Grigorieva}
\affiliation{Institute for Nuclear Research of the RAS,\\
60th October Revolution prospect 7A, Moscow, Russia}

\author{Bohdan Hnatyk}
\affiliation{Astronomical Observatory of Kiev National University, 3
  Observatorna street, 04053 Kiev, Ukraine}

\date{\today}

\begin{abstract}
The dip is a feature in the diffuse spectrum of ultra-high
energy (UHE) protons in the energy range  $1\times 10^{18} - 4\times
10^{19}$ eV, which is caused by electron-positron pair production on
the cosmic microwave background (CMB) radiation. For a power-law
generation spectrum $E^{-2.7}$, the calculated position and shape of
the dip is confirmed with high accuracy by the spectra observed by
the Akeno-AGASA, HiRes, Yakutsk and Fly's Eye detectors. When the
particle energies, measured in these detectors, are calibrated by
the dip, their fluxes agree with a remarkable accuracy. The
predicted shape of the dip is quite robust: it is modified very
weakly when the discreteness and inhomogeneities in the source
distribution are taken into account, and for different regimes of
propagation (from rectilinear to diffusive). The cosmological
evolution of the sources, with parameters inspired by
observations of Active Galactic Nuclei (AGN), also results in the
same shape of the dip. The dip is modified strongly when the
fraction of nuclei heavier than protons is high at injection, which
imposes some restrictions on the mechanisms of acceleration
operating in UHECR sources. The existence of the dip, confirmed by
observations, implies that the transition from galactic to
extragalactic cosmic rays occurs at $E \lsim 1\times 10^{18}$~eV. We
show that at energies lower than a characteristic value $E_{\rm cr}
\approx 1\times 10^{18}$ eV, determined by the equality between the
rate of energy losses due to pair production and adiabatic losses,
the spectrum of extragalactic cosmic rays flattens in all cases of
interest, and it provides a natural transition to a steeper
galactic cosmic ray spectrum. This  transition occurs at some
energy below $E_{\rm cr}$, corresponding to the position of the
so-called second knee. We discuss extensively the constraints on
this model imposed by current knowledge of acceleration processes
and sources of UHECR and compare it with the traditional model of
transition at the ankle.
\end{abstract}

\vskip0.8cm
\maketitle

\section{Introduction}
\label{introduction}
A very important step toward unveiling the origin of the sources of
UHECR is to identify the range of energies where cosmic rays become
mainly of extragalactic origin. For such extragalactic cosmic rays,
in the hypothesis of a proton dominated spectrum,
the propagation in the intergalactic medium induces three features in
the spectrum observed at the Earth: 1) the GZK feature \cite{GZK},
a suppression of the flux at energies in excess of $\sim 10^{20}$
eV, due to the photopion production interactions of cosmic rays
off the CMB photons; 2) a bump (\cite{HS85} - \cite{Stanev00}),
due to the accumulation of particles below the kinematic threshold
for photopion production. As was shown in \cite{BG88}, while the
bump is present in the calculated spectra of single sources, it almost
disappears
in the diffuse spectrum, because the position of the bump depends
upon the source redshift; 3) A dip (\cite{HS85} - \cite{BGG3}),
generated due to pair production, $p+\gamma_{\rm CMB}
\to p+e^++e^-$, where the target is provided by the CMB photons.

The detection of these features would be a definitive test of the
extragalactic origin of UHECR and of the fact that they are mainly
protons. Since the detection of the GZK feature requires very large
statistics of events and, as stressed above, the bump is almost
absent in the diffuse spectrum, at present the spectral feature that
can be detected more easily is the dip. As we show here (see also
\cite{BGG3} - \cite{BGG}), the dip (see Fig.~\ref{fig:mfactor}) is a
quite robust prediction of the calculation and we claim that in fact
it might have already been observed by the AGASA, Fly's Eye, HiRes
and Yakutsk experiments (see \cite{agasa} - \cite{NW} for the data).
However, the detectability of the dip as a feature of the
propagation of cosmic rays on cosmological scales would also imply
that the transition from galactic to extragalactic cosmic rays
should not take place at the {\it ankle}, as has been postulated
since the end of '70s, when this feature was discovered in the
Haverah Park data (see \cite{ankle} for the recent works).

The traditional explanation of the transition from galactic to
extragalactic cosmic rays invokes the intersection between a steep
($E^{-3.1}$) galactic spectrum and a flat ($E^{-\alpha}$ with
$\alpha=2-2.3$) extragalactic spectrum, at the {\it ankle},
located at an energy $E_a \approx 10^{19}$ eV and identified as a
flattening of the spectrum in the data of AGASA, HiRes and Yakutsk
detectors (see Fig. \ref{fig:dips} and \cite{DeMSt} for a general
discussion of the transition).

It is important to stress that in the dip scenario the predicted
spectrum flattens below and above the dip location (see
Fig~\ref{fig:mfactor}). The high energy flattening, at $E_a \approx
1\times 10^{19}$ eV, reproduces perfectly well the observed {\it
ankle}. The low energy flattening, at $E_{\rm cr} \approx 1\times
10^{18}$ eV, obtained for both cases of rectilinear \cite{BGG} and
diffusive propagation \cite{AB,Lem,AB1}, provides the transition to
a steeper galactic component. Note that this property, the
intersection of steep and flat spectral components, is the same for
both models of transition.

We demonstrate here (see also \cite{AB1}) that $E_{\rm cr}$ is
connected with the energy scale $E_{\rm eq} = 2.3 \times 10^{18}$
eV, where the rates of pair production and adiabatic energy losses
are equal.  The visible transition from galactic to extragalactic
cosmic rays occurs at $E_{\rm tr} < E_{\rm cr}$, and this energy
coincides with the position of the {\em second knee}  (Akeno -
$6\times 10^{17}$~eV, Fly's Eye - $4\times 10^{17}$~eV, HiRes -
$7\times 10^{17}$~eV and Yakutsk - $8\times 10^{17}$~eV). The
transition at the second knee was also proposed as a consequence of
the study of the propagation of galactic cosmic rays
(\cite{Biermann} - \cite{Dermer}).

The energy region around
$E_{\rm cr} \approx 10^{18}$ is also expected to correspond to a change
in the chemical composition, from a heavy galactic component to a
proton-dominated composition of UHECR. While HiRes \cite{mass-Hires},
HiRes-MIA \cite{Hi-Mia} and Yakutsk \cite{Glushkov00} data support
this prediction and Haverah-Park \cite{HP} data do not contradict it
at $E \gsim (1 - 2)\times 10^{18}$~eV, the Akeno \cite{mass-Akeno}
and Fly's Eye \cite{FE} data favor a mixed composition, dominated
by heavy nuclei (for a review see \cite{NW} and \cite{Watson04}).

In this paper we shall use the number density of UHECR sources estimated from
the small-scale clustering in the angular distribution of the arriving
particles \cite{ssc}. In \cite{ssa1,ssa2} it was found that
the observed small-scale clustering implies, in the case of rectilinear
propagation of particles, a spatial density of UHECR
$n_s \approx (1 - 3)\times 10^{-5}~{\rm Mpc}^{-3}$. Approximately
the same space density was found from a study of the small-scale clustering
for propagation in magnetic fields \cite{Sato}. This density is of the
same order of magnitude as the density of powerful AGN. 
It is however worth stressing that the simulated AGASA spectra with
this source density are incompatible with the observed AGASA spectrum
at the $5\sigma$ level \cite{danny1}. 
The potential of future detectors such as the Pierre
Auger Observatory to measure the source density from small scale
anisotropies has been discussed in \cite{danny2}.
From the phenomenological point of view,
there are some observational indications of AGN as the sources of
UHECR \cite{TT}, although the subject is still matter of much debate.

The paper is organized as follows: in Section \ref{dip} we discuss
the physics behind the formation of the dip, and compare our
predictions with the data of several experiments. We discuss also
the robustness of the prediction of the dip and some physical
phenomena which modify the shape of the dip. In
Section \ref{transition} we address the more specific issue of the
transition from galactic to extragalactic cosmic rays, stressing the
differences between the {\it dip scenario} and the {\it ankle
 scenario}. In the Appendix we discuss the problems of acceleration relevant
for the dip scenario.  We conclude in Section \ref{conclusions}.

\section{The dip}
\label{dip}

In this section we describe in detail the physical arguments that
explain the formation of the dip in the spectrum.
\label{sec:dip}
\begin{figure}[ht]
  \begin{center}
    \includegraphics[width=8.0cm]{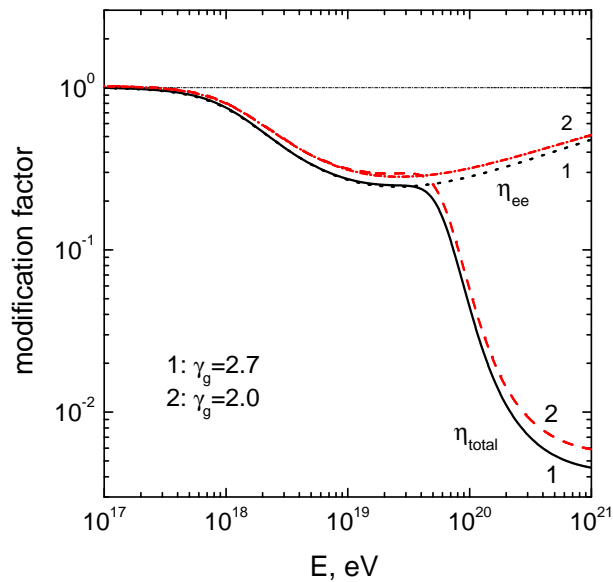}
\end{center}
  \caption{Modification factor for a power-law generation spectrum
with slope $\gamma_g=2.0$ and 2.7. The horizontal line $\eta=1$ corresponds to
adiabatic energy losses only. The curves $\eta_{ee}$ and $\eta_{tot}$
correspond respectively to the modification factor for adiabatic and pair
production energy losses and the modification factor where all losses
are taken into account.}
  \label{fig:mfactor}
\end{figure}

In order to do this, we use the formalism of the {\it modification
factor}, first introduced in \cite{BG88} and defined as the ratio of
the spectrum $J_p(E)$ with all energy losses taken into account, and
the unmodified spectrum $J_p^{\rm unm}$, where only adiabatic energy
losses (red shift) are included: $\eta(E)=J_p(E)/J_p^{\rm unm}(E)$.
The spectrum $J_p(E)$ can be calculated from the conservation of the
number density of particles as
\beq
n_p(E,t_0)dE= \int_{t_{min}}^{t_0} dt Q_{\rm gen}(E_g,t)dE_g,
\label{conserv}
\eeq
where $n_p(E,t_0)$ is the space density of UHE protons at the
present time, $t_0$, $Q_{\rm gen}(E_g,t)$ is the generation rate per
comoving volume at cosmological time t, and $E_g(E,t)$ is the generation
energy at time t for a proton with energy E at $t=t_0$. This energy
is found from the loss equation $dE/dt=- b(E,t)$, where $b(E,t)$ is
the rate of energy losses at epoch t. The spectrum, Eq. (\ref{conserv}),
calculated for a power-law generation spectrum $\propto E^{-\gamma_g}$
and for a homogeneous distribution of sources, is called
{\em universal spectrum}. The important feature of the universal spectrum
is its independence of the mode of propagation: it is the same for
rectilinear propagation and propagation in arbitrary magnetic fields. This
property of the universal spectrum is guaranteed by the propagation theorem
\cite{AB}, according to which the spectra do not depend on the
propagation mode if the distance between sources is less than any
propagation length, e.g. energy attenuation or diffusion length.
For homogeneous distribution of the sources with vanishing distance
between them the propagation theorem is obviously fulfilled.

The generation rate $Q_{\rm gen}(E,t)$ might include the cosmological
evolution of the sources. In the results presented in this section,
we shall not include it in the calculations for two reasons:
(i) The evolution involves at least two free parameters, $m$ and
$z_{\rm max}$, where $m$ is the exponent in the evolution rate
$(1+z)^m$, and $z_{\rm max}$ is the maximum redshift
up to which evolution takes place. This makes the fit to the data
more arbitrary. (ii) Evolution is a very model-dependent phenomenon, and
as such we will discuss it later in Section~ \ref{evolution}, regarding
it as an uncertainty in the predictions.

Since the injection spectrum $E^{-\gamma_g}$ enters both the numerator
and the denominator of $\eta (E)$, one may expect that the
modification factor depends weakly on $\gamma_g$ and numerical
calculations confirm it.

In Fig. \ref{fig:mfactor} we plot the modification factor as a function
of energy for two slopes of the injection spectrum, $\gamma_g=2.0$
and $\gamma_g=2.7$. As expected, the differences are quite small.

\begin{figure}[ht]
\begin{minipage}[h]{8cm}
\centering
\includegraphics[width=7.6cm,clip]{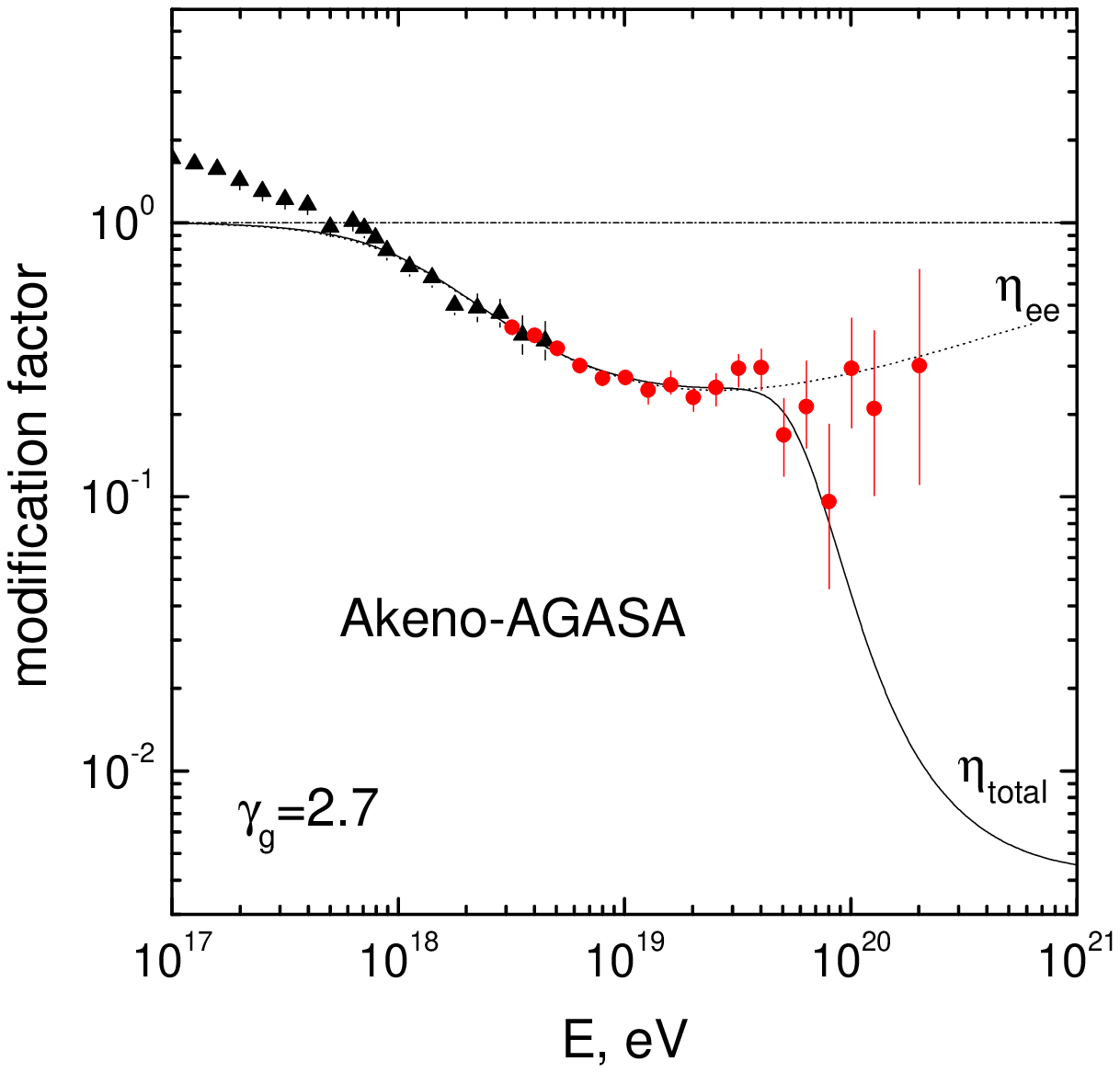}
\end{minipage}
\hspace{2mm}
\begin{minipage}[h]{8cm}
\centering
\includegraphics[width=7.6cm,clip]{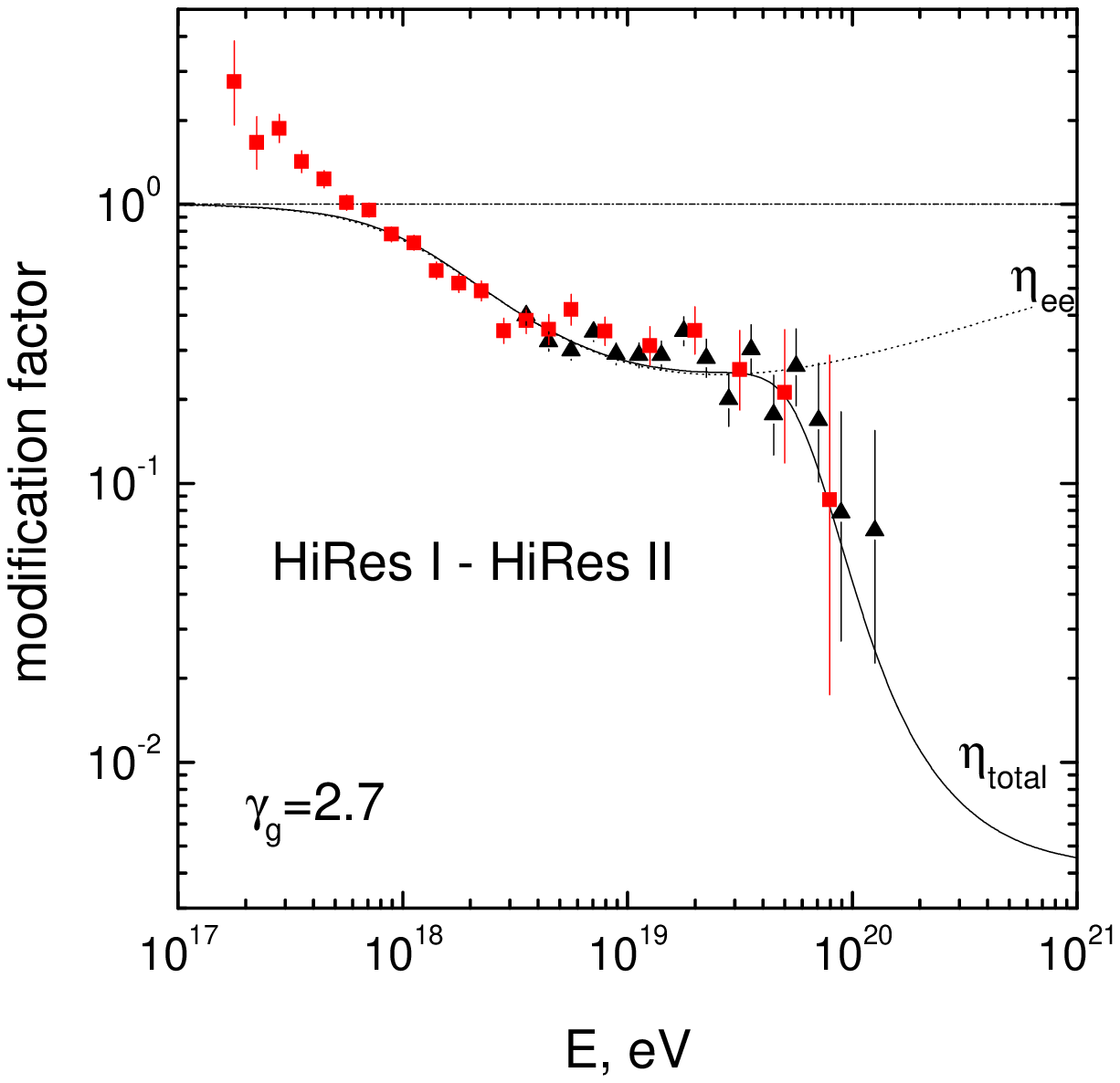}
\end{minipage}
\vspace{2mm}
\begin{minipage}{8cm}
\centering
\includegraphics[width=7.6cm,clip]{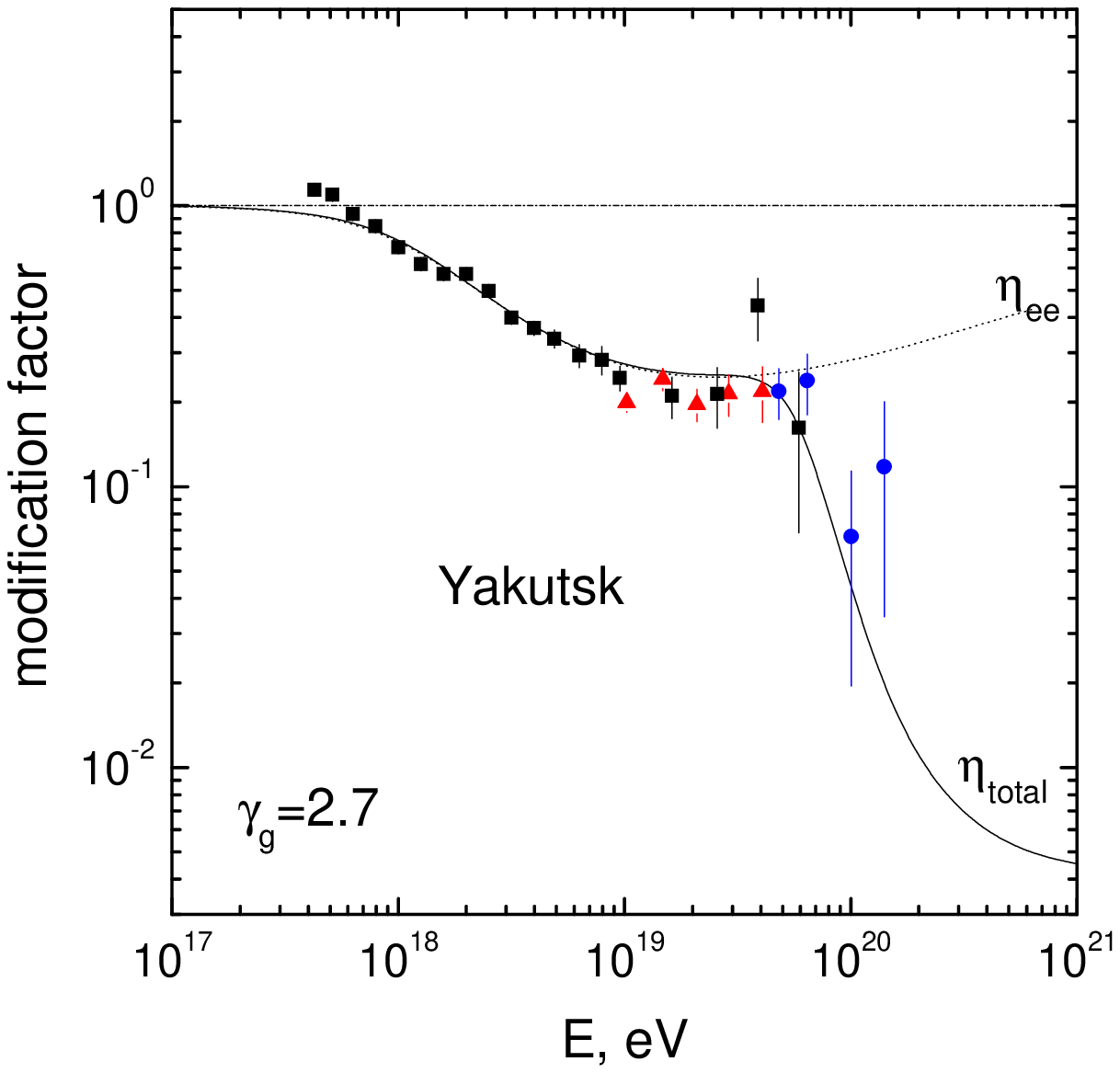}
\end{minipage}
\hspace{7mm}
\begin{minipage}[h]{8cm}
\centering
\includegraphics[width=7.6cm,clip]{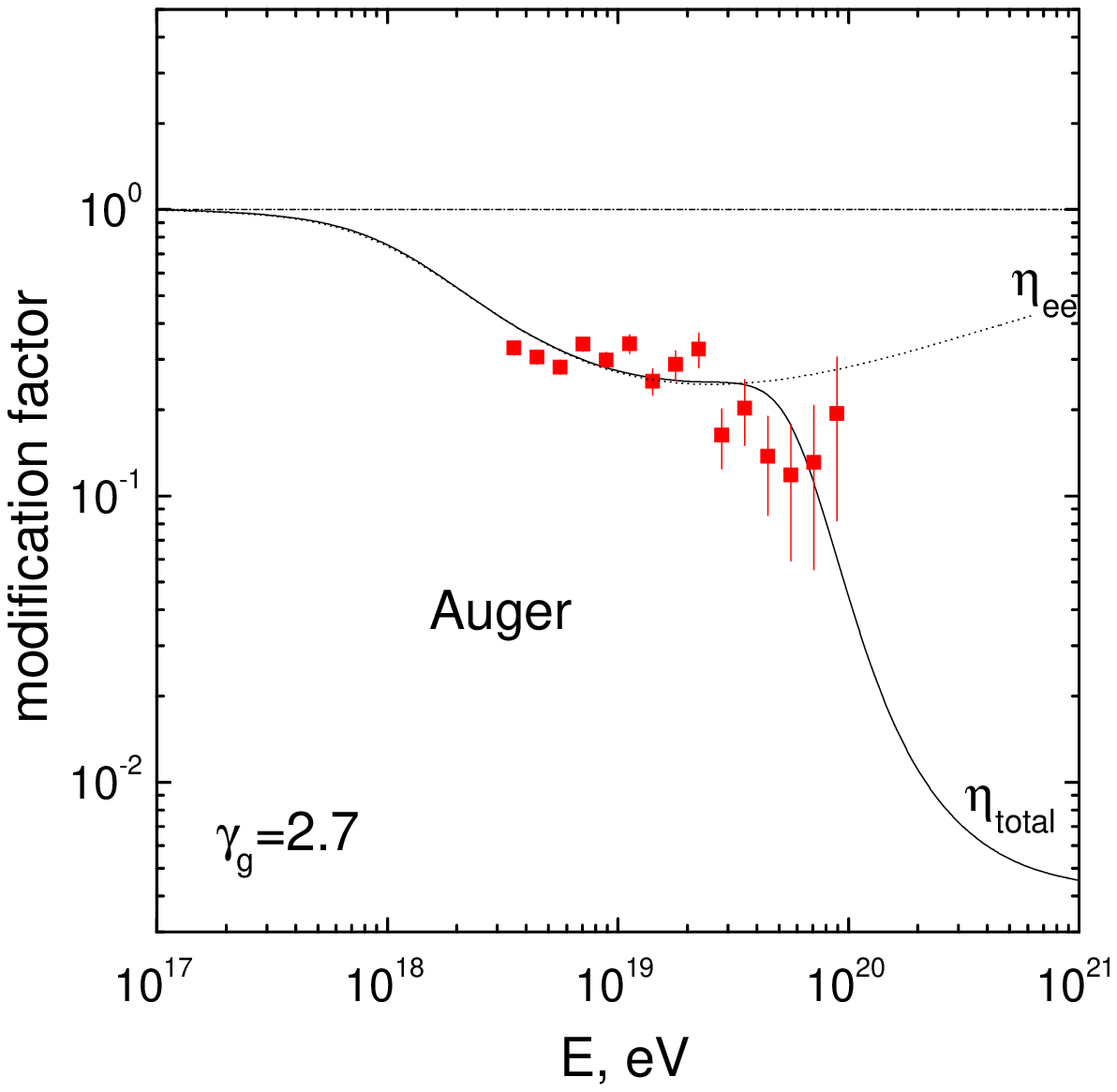}
\end{minipage}
\caption{\label{fig:dips} Predicted dip in comparison with AGASA, HiRes,
Yakutsk and Auger\protect\cite{Auger} data.}
\end{figure}

In Fig. \ref{fig:dips} we show the comparison of the modification
factor calculated for $\gamma_g=2.7$  with the observational
data of AGASA, HiRes, Yakutsk and Auger. The dip, i.e.
the modification factor $\eta_{ee}(E)$, is well confirmed by the data at
energy below $E \approx 4\times 10^{19}$~eV, above which the photopion
production dominates (see Fig.~\ref{fig:dips}). Fly's Eye data, not
shown here, confirm the dip equally well. Auger spectrum does not
contradict the high energy part of the dip, but needs continuation of
the spectrum to lower energies to test the dip as a whole.

At energy $E\geq 1\times 10^{19}$~eV the dip shows a flattening, which
explains the ankle, seen in the data in Fig.~\ref{fig:dips} at this
energy.

By definition the modification factor must be less than unity.
At energy $E < 1\times 10^{18}$~eV the modification factors of
AGASA-Akeno and HiRes exceed this bound. This signals the appearance
of another component, which is most probably given by galactic cosmic
rays. This is the first indication in favor of a transition from
extragalactic to galactic cosmic rays at $E \sim 1\times 10^{18}$~eV.

The best fit to the data provided by analytical calculations
corresponds to $\gamma_g=2.7$, though $2.6 \leq \gamma_g \leq 2.8$
provide an acceptable description of the data. The detailed
Monte-Carlo simulations of the spectra at $E \geq 3\times 10^{18}$~eV,
accounting for statistical errors in the energy determination of the
events, lead to a best fit injection spectrum with slope
$\gamma_g=2.6$ \cite{DBO}, in rather good agreement with the results
of analytical calculations. In addition to the statistical errors, the
simulations in \cite{DBO} may also account for a systematic error in
the energy determination. For most currently operating experiments
such error is of order 20\% and sometimes in excess of this.
The Monte-Carlo simulations of \cite{DBO} lead to the
conclusion that the alleged discrepancy between the AGASA
and HiRes experiments could be explained in terms of a combination of
statistical, systematic errors in the energy of the events, and
statistically limited number of events at energies above $\sim 10^{20}$
eV. This conclusion was strengthened in \cite{DBO1} where the same
authors showed that the realizations of the simulated propagation of UHECR
that are found to have 11 or more events at energy above $10^{20}$ eV
resemble very closely the AGASA data, although the {\it average}
spectrum has a pronounced GZK feature. In the same paper, the
authors also make an attempt to extract events at random directly from
the AGASA data and calculate the probability of obtaining the HiRes
spectrum by chance. In both cases the alleged discrepancy between the
two experiments is found to be statistically not very significant.

\begin{figure}[ht]
  \begin{center}
    \includegraphics[width=17.0cm]{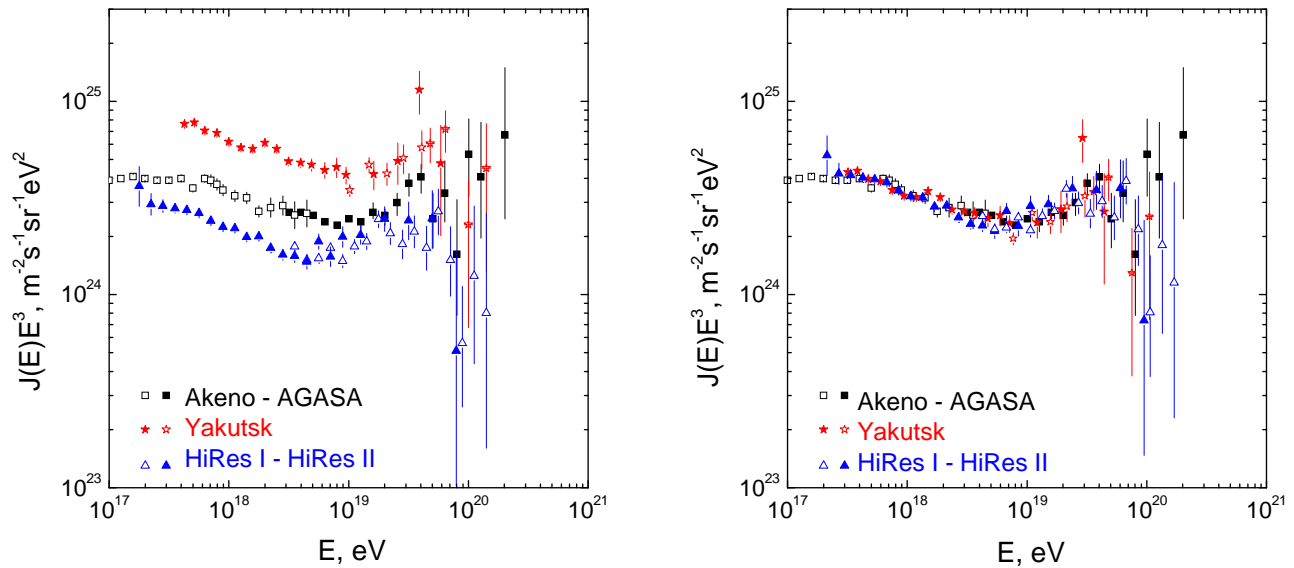}
\end{center}
  \caption{The spectra as measured by  Akeno-AGASA, HiRes and Yakutsk
arrays (left panel) and after energy calibration by the dip (right panel).
}
  \label{fig:AgHiYa}
\end{figure}

The shape of the dip can be used for the energy calibration of the
detectors. Shifting the energies by a factor $\lambda$ for each detector
in the energy interval of the dip ($(1 - 40)\times 10^{18}$~eV) one
can determine the minimum $\chi^2$ and from it deduce the value of
$\lambda$ that best fits the data. This procedure leads to
$\lambda_{\rm Ag}=0.9$, $\lambda_{\rm Hi}=1.2$ and $\lambda_{\rm
  Ya}=0.75$ for AGASA, HiRes and Yakutsk detector, respectively. 
It is worth noticing that the required correction factor is less than
unity for ground arrays and exceeds unity for fluorescence experiments. 
It is impressive that after this shift (calibration by the dip) the spectra
of these three detectors agree very well (see Fig.~\ref{fig:AgHiYa}).
The results shown in Fig.~\ref{fig:AgHiYa} have already been presented
by some of us at different conferences starting from 2004 (see
e.g. \cite{Berez05}).
Recently AGASA collaboration \cite{Teshima06} has reduced the energies
by about 10\% as we predicted.


\subsection{Robustness and caveats}
\label{robust}

The prediction that a dip is present in the spectrum of extragalactic
cosmic rays has been obtained using what we called the universal
spectrum. Several assumptions have been used in the calculation (see
below) and we need to assess what could be their role in changing the
basic predictions of the dip scenario. We will prove that within some
reasonable limitations, the universality of the spectrum is not
substantially changed, in particular in the region of energy around
the dip, $1\times 10^{18}~{\rm eV} \leq E \leq 4\times 10^{19}$~eV.
The GZK feature, located at higher energies is expected to exhibit
noticeable deviations from the universal spectrum due to possible local
inhomogeneities in the source distribution (local overdensity or
deficit of the sources \cite{Blanton,BGG}), due to a possible acceleration
related cutoff and discreteness in the source distribution \cite{BGG}.

We start with listing the phenomena which may modify the universal
spectrum:
\begin{itemize}
\item[1)] Discreteness  in the source distribution for the case
  of  propagation in magnetic fields;
\item[2)] Inhomogeneous source distribution;
\item[3)] Cosmological evolution of the sources;
\item[4)] Energetics corresponding to the best fit injection spectrum;
\item[5)] Chemical composition.
\end{itemize}

\subsubsection{Discreteness in the source distribution for propagation
 in  magnetic fields}
\label{discrete}

Here we discuss how the discreteness in the source distribution affects
the universal spectrum. We distinguish the weak magnetic field case,
when protons propagate with moderate deflections, and the case of
strong magnetic field when propagation is diffusive.

\begin{figure}[ht]
  \begin{center}
    \includegraphics[width=10.0cm]{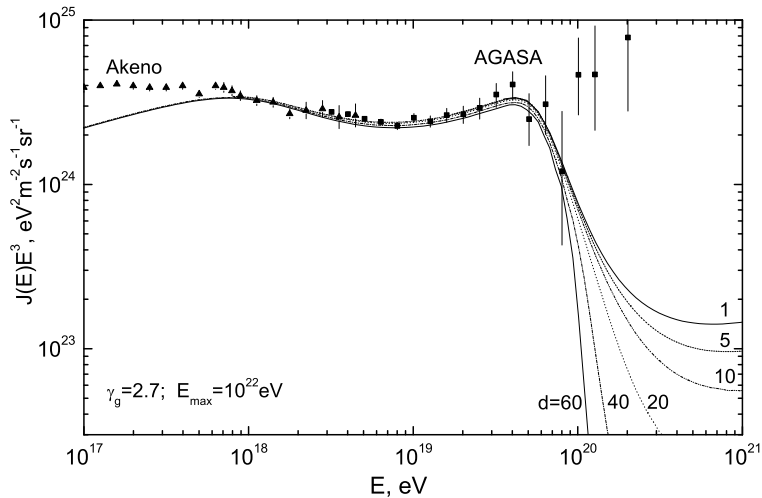}
\end{center}
  \caption{UHE proton spectra for rectilinear propagation from discrete
    sources, located in the vertexes of the cubic grid with spacing
    d=60, 40, 20, 10, 5 and 1~Mpc. The calculations are performed
for $z_{\rm max}=4$, $E_{\rm max}=1\times 10^{22}$~eV and $\gamma_g=2.7$.
}
  \label{fig:discrete}
\end{figure}

We shall start with the unrealistic case of a rectilinear propagation of
protons in the dip energy region, which is formally valid in the
absence of magnetic field. In Fig.~\ref{fig:discrete} the spectra are
calculated in the case of source distributions with different
distances between them, as indicated in the plot.
One can see that while the shape of the GZK steepening is noticeably
modified as the distance $d$ between sources changes, the dip is very weakly
affected by it, and the spectrum remains universal (the curve with
d=1~Mpc in Fig.~\ref{fig:discrete}). This result directly follows
from the propagation theorem, because the only propagation length scale
in this problem, the energy attenuation length, exceeds 1000 Mpc in
the region of the dip, being thus considerably larger than the
distance between the sources used in the calculations.

It is clear that with a weak magnetic field the spectra remain universal,
like for rectilinear propagation, if particles are not deflected by large
angles, when propagating from a source to the observer. It is less trivial
that the spectra shown in Fig.~\ref{fig:discrete} should coincide with
the spectra obtained for the tangled trajectories (including
diffusion) for the same $d$ as in the case of rectilinear
propagation. This follows again from the propagation theorem which
states that for diffusive propagation the spectrum remains universal
provided that the attenuation length and the diffusion length are
larger than the mean separation $d$ between sources.

\begin{figure}[ht]
  \begin{center}
    \includegraphics[width=10.0cm]{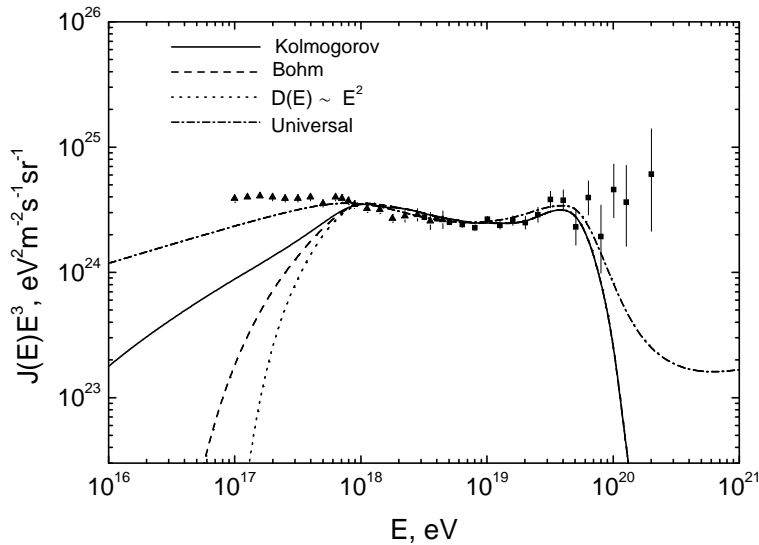}
\end{center}
  \caption{Diffusive energy spectrum for $B_c=1$~nG, $l_c=1$~Mpc,
$d=50$~Mpc  and $\gamma_g=2.7$ for the Bohm, Kolmogorov and
$D(E)\propto E^2$ diffusion. The dash-dotted curve shows the
universal spectrum. The data of Akeno-AGASA are also shown.  }
  \label{fig:diffuse}
\end{figure}

As pointed out in \cite{AB1}, at energy $\sim 1\times 10^{18}$ eV the
maximum distance $R_{\rm max}(E)$ from which particles can reach the
observer suffers a sharp increase (the so-called antiGZK effect
\cite{AB1}). The energy at which this takes place was found to be
independent of the choice of the diffusion coefficient
\cite{AB1}. This is illustrated in Fig. \ref{fig:diffuse} where it is
visible that for different choices of the diffusion coefficient
(Kolmogorov, Bohm and $D(E)\propto E^2$) the curves depart from
the universal spectrum at the same energy $E_{\rm cr} =1\times 10^{18}$~eV.
For rectilinear propagation (Fig.~\ref{fig:discrete}) one may see the
same value of $E_{\rm cr}$.

This universality of $E_{\rm cr}$ value  may be explained recalling
the proximity of
$E_{\rm cr}$ to the energy $E_{\rm eq}$, where pair-production
energy losses are as fast as the adiabatic energy losses.
In terms of evolution of the generation energy with the look-back
time, when the proton energy reaches $E_{\rm eq}$, its energy
increases fast, the diffusion length grows even faster and
$R_{\rm max}(E_{\rm cr})$ becomes large in an almost discontinuous way
(see \cite{AB1} for analytical and numerical calculations).
In a semi-quantitative way, the connection between $E_{\rm cr}$
and $E_{\rm eq}$ can be expressed as
$E_{\rm cr}=E_{\rm eq}/(1+z_{\rm eff})^2$, where $z_{\rm eff}$ is an
effective redshift of the sources
contributing to the flux of cosmic rays at energy $\sim E_{\rm cr}$.
A simplified analytical estimate for $\gamma_g=2.6 - 2.8$ gives
$1+z_{\rm eff} \approx 1.5$ and hence $E_{\rm cr} \approx 1\times 10^{18}$ eV.

While the position $E_{\rm cr}$ and the high energy behavior of the
'cutoff' (in fact in terms of $J(E)$ this appears as a flattening of
the spectrum) are determined by an increase of energy losses, there
is a more evident spectral steepening and cutoff at lower energies,
caused by the magnetic horizon. In a very rough approximation,
particles cannot reach the observer from distances larger than
$R_{\rm hor}(E) \sim \sqrt{D(E)t_0}$. When $E$ is small enough, so
that $R_{\rm hor}(E)$ is smaller than the distance $d$ to the nearby
sources, the spectrum develops an exponential cutoff. At low
energies, where energy losses of protons are only adiabatic and
diffusion coefficient is $D(E) \propto E^{\alpha}$, the magnetic
horizon can be calculated (see \cite{AB1}) as \beq R_{\rm
hor}(E)=2\left (\frac{D(E)}{\alpha H_0}\right )^{1/2} \left
(e^{\alpha H_0 t_0}-1 \right )^{1/2}, \label{mhorizon} \eeq where
$H_0$ is the Hubble parameter and $t_0$ is the age of the universe.
For $B_c=1$~nG and $l_c=1$~Mpc one finds the energy of the magnetic
horizon cutoff from the condition $R_{\rm hor}(E_{\rm cut})=d$,
where $d$ is the distance between sources, as $E_{\rm cut}=2.3\times
10^{15}$~eV for $d=30$~Mpc and  $E_{\rm cut}=4.8\times 10^{16}$~eV
for $d=50$~Mpc, both for the Kolmogorov diffusion. For the case of
Bohm diffusion these energies are larger.

We present in Fig.~\ref{fig:diffuse} the  spectra calculated for
diffusive propagation in relatively strong turbulent magnetic field with basic
turbulent length $l_c= 1$~Mpc, with coherent magnetic field on this
scale $B_c= 1$~nG and for separation between sources $d=50$~Mpc. At
energies higher than $E_{\rm cr}$ diffusion proceeds in the regime
where $D(E) \propto E^2$, (see \cite{AB1}), while at lower energies
we assume the appropriate diffusion regimes with $D(E) \propto E^{\alpha}$,
as indicated in
Fig.~\ref{fig:diffuse}. One can see that the dip calculated in the
diffusive approximation differs very little from the one in the
universal spectrum, shown by dash-dot line.

We conclude this section asserting that the presence of magnetic
fields in the universe modify quite weakly, within a wide range of
parameters, the shape of the dip calculated in the form of a
universal spectrum. We provided evidences that a transition from
extragalactic to galactic cosmic rays occurs at energy $E_{\rm
cr}=1\times 10^{18}$~eV, independently of the mode of propagation
(from rectilinear to diffusive). However, for any reasonable
extragalactic magnetic fields the propagation of protons and nuclei
at  $E < 1\times 10^{18}$~eV is expected to be diffusive. Even for
random magnetic fields with parameters $l_c= 1$~Mpc and $B_c=
0.1$~nG, taken as average over the universe, the diffusion length at
$E=3\times 10^{17}$~eV, is only $\sim 10$~Mpc, which is considerably
smaller than the size of the region contributing the observed
diffuse flux at this energy. This makes the low-energy diffusive
cutoff, which provides the transition in our model, most reliable.


\subsubsection{Inhomogeneity in the source distribution}
\label{inhomog}

It is often argued that the distribution of matter in the universe is
not homogeneous and that the inhomogeneous distribution of the sources
of UHECR may have an effect on the observed spectrum. This comment
clearly applies to all inhomogeneities on scales smaller than the loss
length of particles with given energy. For instance, at extremely high
energies (around $10^{20}$ eV) the loss length is $20 - 100$~Mpc.
On these scales the universe is indeed inhomogeneous, and this
may result in a shape of the GZK feature which reflects this
inhomogeneity: a local overdensity (deficit) would make the GZK
feature less (more) pronounced.

On the other hand, we know from cosmological observations that the
universe is homogeneous and isotropic on the scale of the cosmological
horizon, and in fact already on scales of the order of $l\gsim 100$ Mpc.
The particles from the dip, i.e. with energies below $4\times 10^{19}$~eV,
have attenuation length of the order of 1000~Mpc, and thus the
dip shape is insensitive to inhomogeneities, provided they occur
on spatial scales smaller than $\sim 1000$~ Mpc. We checked this
conclusion by including in the calculations some spatial
inhomogeneities on scales of 100~Mpc.


\subsubsection{Cosmological evolution of UHECR sources}

The cosmological evolution of the sources, namely the increase in the
source luminosity or comoving density with red-shift $z$, is
observed for many classes of astrophysical objects. The evolution
is reliably observed for the star formation rate in normal galaxies,
but this case is irrelevant for most of the cases of our concern,
because neither stars nor normal galaxies can be the sources of
UHECR due to insufficient cosmic-ray luminosities $L_p$ and maximum
energy at acceleration $E_{\rm max}$. An  exception to this
rule might be represented by the case of Gamma Ray Bursts (GRBs).
\label{evolution}
\begin{figure}[ht]
  \begin{center}
    \includegraphics[width=10.0cm]{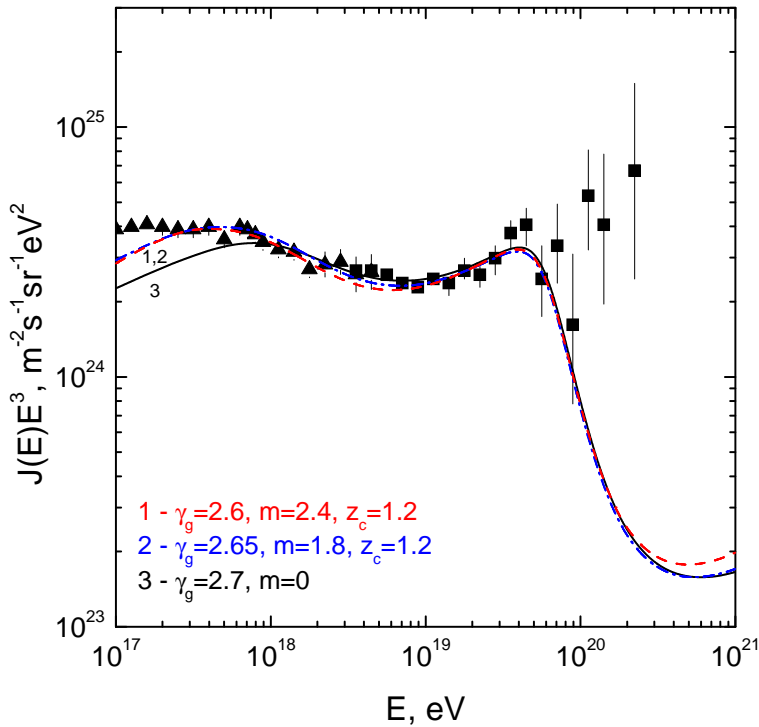}
\end{center}
  \caption{Dip calculated in the models with cosmological evolution.
The parameters of evolution used in the calculations for curves 1 and
2  are close to those observed for AGN. The curve 3 is the universal
spectrum with $m=0$.}
  \label{fig:evolution}
\end{figure}

Active Galactic Nuclei (AGN), which are most probable candidates for
UHECR sources, exhibit evolution in the radio, optical
and X-ray bands. X-rays are probably the most relevant tracer for
evolution of AGN as  the sources of UHECR: X-rays
are produced in accretion disks around massive black holes, and the
X-ray luminosity is connected thus with the accretion rate. UHECR
are probably also connected with the accretion power of massive
black holes through the production of jets and generation
of shock waves in the jets and radio lobes. According to
recent detailed analysis \cite{Ueda,Barger} the
evolution of AGN seen in X-ray radiation can be described in terms of
the factor $(1+z)^m$ up to $z_c \approx 1.2$ and is saturated at larger
$z$.  In \cite{Ueda} the pure luminosity evolution and pure density
evolution is allowed with $m=2.7$ and $m=4.2$, respectively and
with $z_c \approx 1.2$ for both cases. In \cite{Barger} the
pure luminosity evolution is considered as preferable with $m=3.2$
and $z_c =1.2$. These authors do not distinguish between
different morphological types of AGN. It is possible that some AGN
undergo weak cosmological evolution, or no evolution at all. For
instance BL Lacs, which are suspected as sources of observed UHECR
\cite{TT}, show {\em negative} evolution ($m<0$) \cite{Morris}, which
should be most probably interpreted as weak or absent evolution.

The effect of the evolution of the sources of UHECR on
the shape of the dip was discussed in \cite{BGG} and in a recent
review \cite{blasirev}.
In the case of UHECR there is no need to distinguish between luminosity
and density evolution, because the diffuse flux is determined by the
comoving energy-density production rate ({\em emissivity})
${\mathcal L}=L_p n_s$ , where $L_p$ is the cosmic ray luminosity and
$n_s$ is the space density of the sources.

In Fig.~\ref{fig:evolution} we present the calculated spectrum for
evolutionary models, inspired by the data cited above. For
comparison we show also the case of absence of evolution $m=0$. As
our calculations show, in most cases the negative evolution ($m
<0$) results in the same shape of the dip as the no-evolution case
$m=0$.

The universal spectra, obtained for sources evolving up to $z_c > 1$,
fit the observational data down to $E \sim 3\times 10^{17}$~eV and
even at lower energies. However, for reasonable magnetic fields in the
intergalactic medium, protons with these energies have small
diffusion lengths and the universal spectrum fails at
$E < 1\times 10^{18}$~eV, exhibiting a diffusion 'cutoff' that starts
at energy $E_{\rm cr}$.

We conclude that at present evolutionary models can fit the shape of
the dip as well as models without evolution ($m=0$).

\subsubsection{Energetics }
\label{energetics}

The universal spectrum, which fits the observed dip, requires an
injection spectrum with a slope $\gamma_g=2.6-2.7$. The normalization
to the observed flux needs the emissivity (energy-density production
rate) at $t=t_0$~~ ${\mathcal L}_0 \propto E_{\rm min}^{-(\gamma_g-2)}$,
where $E_{\rm min}$ is the minimum acceleration energy. For low
$E_{\rm min} \sim 1$~GeV the required emissivity is too high. In order
to prevent this energetic crisis, it was suggested phenomenologically
in Refs.~\cite{BGG} and \cite{BGH} that the generation rate per unit
comoving volume $Q(E_g)$ may have the form
\begin{equation}
Q_{\rm gen}(E_g)=\left\{ \begin{array}{ll}
\propto E_g^{-2}                       ~&{\rm at}~~ E_g \leq E_c\\
\propto E_g^{-2.7}                     ~&{\rm at}~~ E_g \geq E_c,
\end{array}
\right.
\label{broken}
\end{equation}
where the spectrum $\propto E^{-2}$ is due to non-relativistic shock
acceleration. Recently, an interesting idea was put forward in
Ref.~\cite{michael}, due to which the broken spectrum Eq. (\ref{broken})
can be realized. The authors of \cite{michael} observed that while
the slope of the acceleration spectrum (e.g. 2.0 for non-relativistic shocks
and 2.2 for relativistic shocks) is universal, the maximum acceleration
energy $E_{\rm max}$ depends on individual characteristics of the sources,
such as magnetic field and/or size. As a result the sources should
be expected to have a distribution of maximum energies at the different
acceleration sites, $n_s(E_{\rm max})$.
This distribution naturally results in a complex spectrum of the form
given in Eq. (\ref{broken}), with $E_c$ being a free parameter.

In section \ref{Emax} we propose another model for the complex
spectrum (Eq. \ref{broken}). It is based on a possible correlation
of $E_{\rm max}$ with source luminosity. The distribution of sources
over luminosities results in the complex spectrum in the form of Eq.
(\ref{broken}) for the generation rate per comoving volume, $Q_{\rm
gen}(E_g)$.

With the complex spectrum Eq.~(\ref{broken}) we obtain for the spectrum
shown in the right panel of Fig.~\ref{fig:AgHiYa}, the emissivity
${\mathcal L}_0=3.7\times 10^{46}$~erg Mpc$^{-3}$ yr$^{-1}$ for
$E_c=1\times 10^{18}$~eV.  Using the sources
space density $n_s=2\times 10^{-5}$~Mpc$^{-3}$, as deduced from
small scale anisotropy \cite{ssa1,ssa2}, we arrive at the cosmic ray
luminosity of a source $L_p=5.9\times 10^{43}$,~$2.6\times 10^{44}$
and $1.2\times 10^{45}$~erg/s for $E_c$ equals to $1\times 10^{18}$~eV,
$1\times 10^{17}$~eV and $1\times 10^{16}$~eV, respectively. These
luminosities fit well the energetics potential of AGN.

It is necessary to emphasize that low $E_c$  in Eq.~(\ref{broken})
does not change our
conclusion about the transition from galactic to extragalactic
cosmic rays, provided $E_c$ is below or close to $1\times 10^{18}$~eV.


\subsubsection{Chemical Composition}
\label{chemical}

\begin{figure}[ht]
\begin{minipage}[h]{8cm}
\centering
\includegraphics[width=7.6cm,clip]{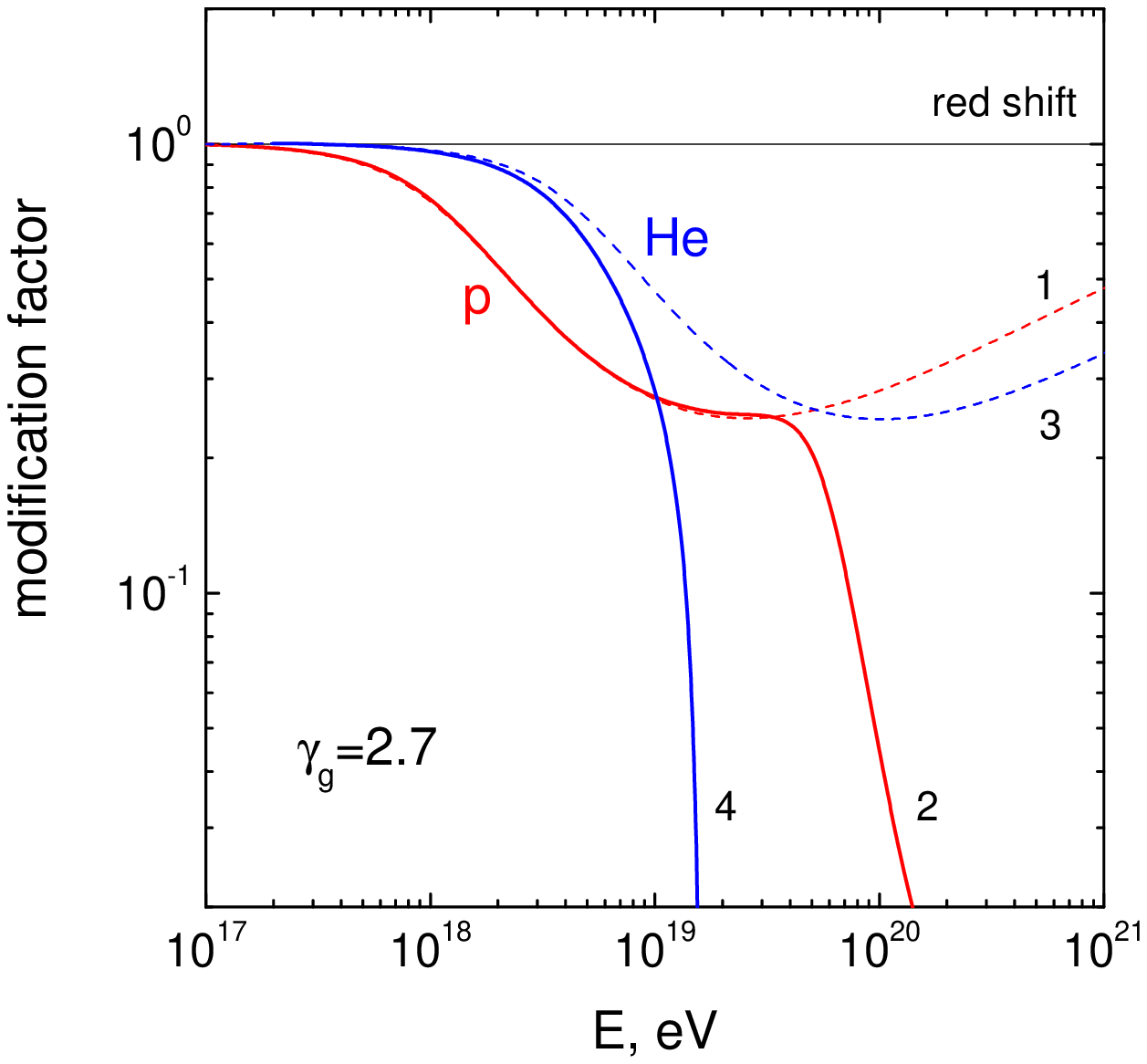}
\end{minipage}
\hspace{5mm}
\begin{minipage}[h]{8cm}
\centering
\includegraphics[width=7.6cm,clip]{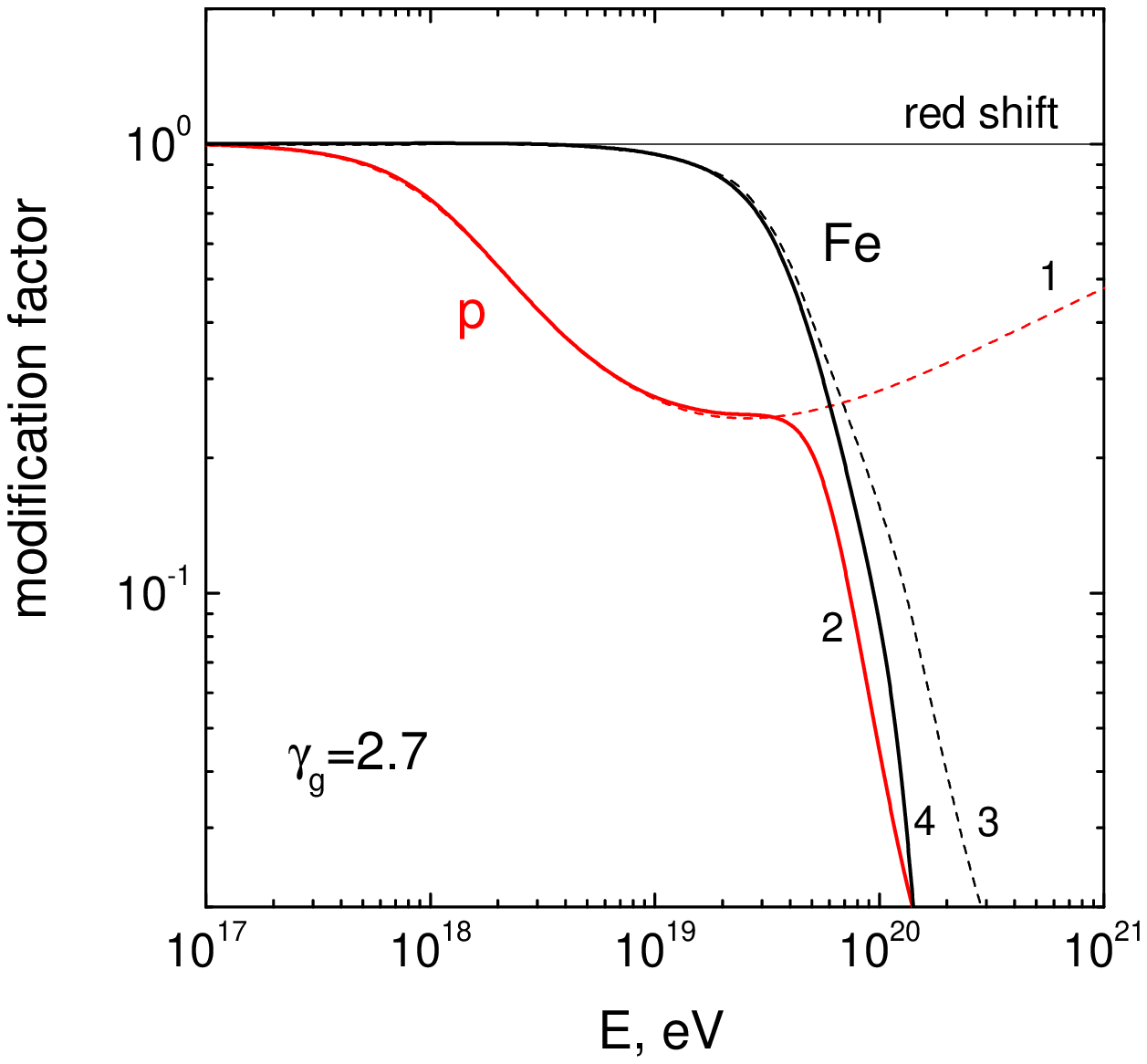}
\end{minipage}
\caption{Modification factors for helium and iron
 nuclei in comparison with that for protons. Proton modification
 factors are given by curves 1 and 2.
 Modification factors for nuclei are shown as curve 3 (adiabatic
 and pair production energy losses) and curve 4 (with
 photodisintegration included).}
\label{fig:dip-nucl}
\end{figure}

The presence of nuclei heavier than protons in the primary UHECR flux can
substantially modify the proton dip \cite{BGG3,Allard} and affect the
agreement with observational data shown in Fig.~\ref{fig:dips}.
In Fig~\ref{fig:dip-nucl} the modification factors for helium and iron
nuclei are presented in comparison with the proton modification factor.
One can see that the presence of 15 - 20 \% of nuclei in the primary flux
affects the good agreement with observations.

The modification factor for a mixed composition can be calculated by
introducing a mixing parameter $\lambda=Q_A^{\rm unm}(E)/Q_p^{\rm
  unm}(E)$ as
\beq
\eta (E)= \frac{\eta_p(E)+\lambda \eta_A(E)}{1+\lambda},
\label{mod-mix}
\eeq
where $Q_A^{\rm unm}(E)$ is the injection spectrum of nuclei with mass
number $A$ at the source.
\begin{figure}[ht]
\begin{minipage}[h]{8cm}
\centering
\includegraphics[width=7.6cm,clip]{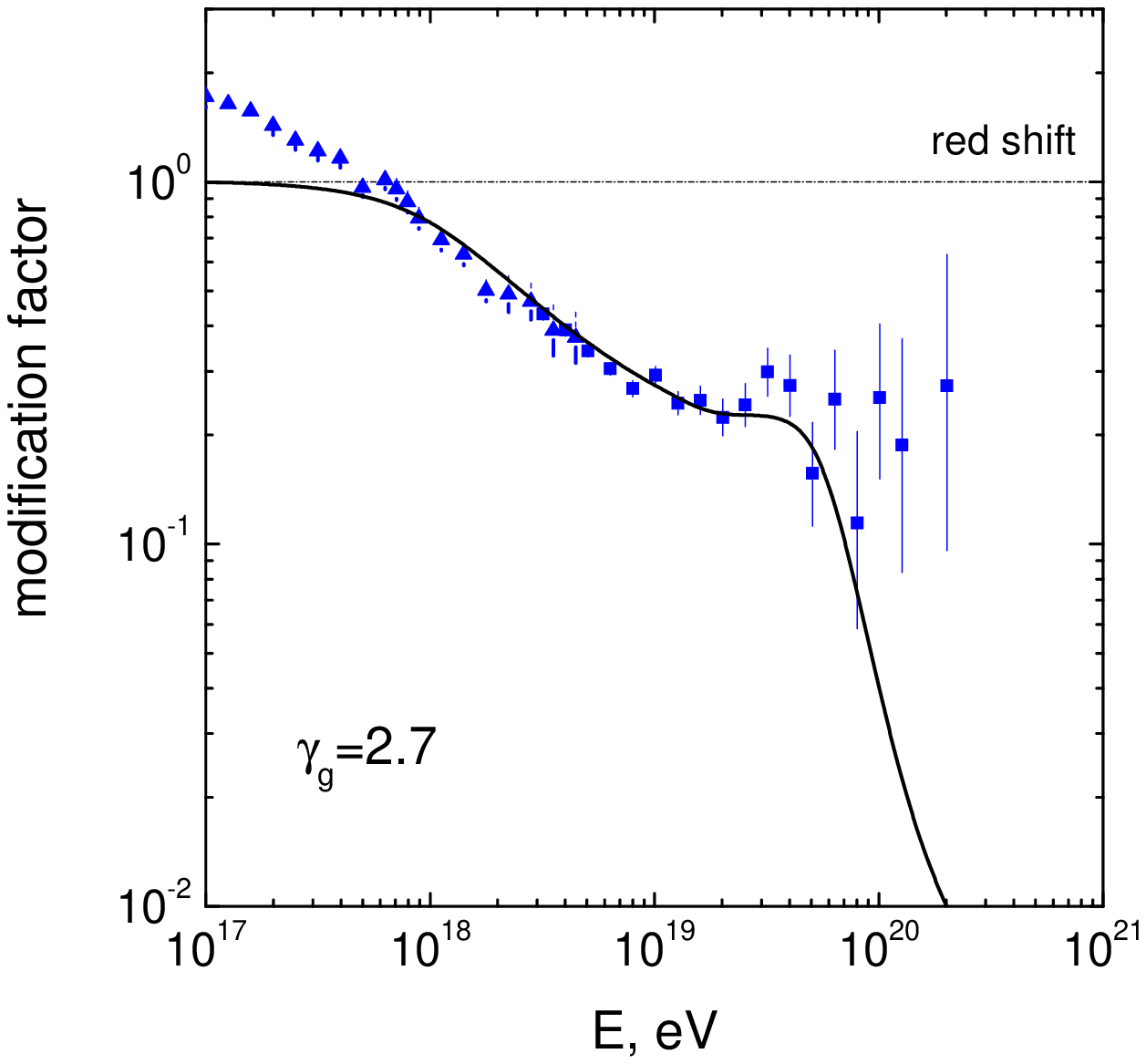}
\end{minipage}
\hspace{5mm}
\begin{minipage}[h]{8cm}
\centering
\includegraphics[width=7.6cm,clip]{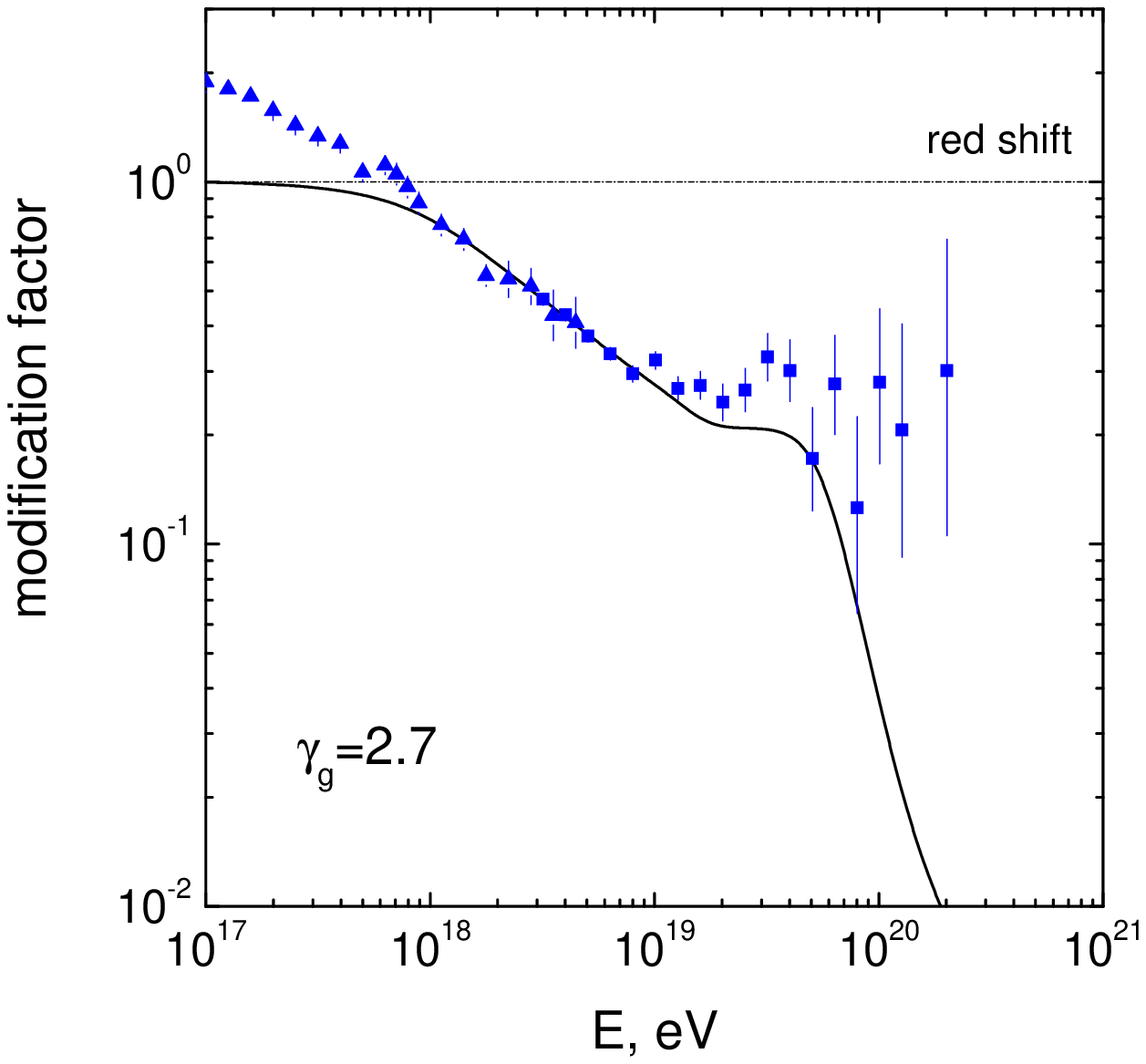}
\end{minipage}
\caption{\label{fig:dip-mix} Modification factors for the mixed composition
of protons and helium nuclei in comparison with the AGASA data. The left
panel corresponds to mixing parameter $\lambda=0.1$, and the right
panel to $\lambda=0.2$.
}
\end{figure}
The mixing parameter $\lambda$ is determined  primarily by the ratio of
the number densities $n_A/n_H$ in the gas where acceleration occurs.
The largest ratio $n_A/n_H$  is given by helium which has basically
a cosmological origin. The cosmological mass fraction of helium
$Y_p=0.24$ results in $n_{\rm He}/n_H=0.079$.
In Fig.~\ref{fig:dip-mix} the modification factors for the mixed
composition of protons and helium with $\lambda=0.1$ and
$\lambda=0.2$ are shown. In the former case the agreement is
sufficiently good, in the latter case the agreement is noticeably
worse than for a pure proton composition.

Apart from the density ratio, the mixing parameter depends on the
details of the acceleration mechanism. In principle, if acceleration
takes place in a relatively cold medium, where helium is not ionized
(the ionization potential for helium is very high: 24.4~eV and
54.4~eV for the first and second ionization potentials,
respectively), helium nuclei may be not accelerated at all. This
possibility can occur for the case of induced electric fields in the
plasma. However, for shock acceleration the fraction of accelerated
nuclei is determined by their injection into the shock acceleration
region (see e.g. \cite{berez}), and even for low temperature of the
upstream gas, the rate of injection of nuclei in the downstream
region can be high. For non-relativistic shocks the ratio of
temperatures downstream and upstream can be low if the Mach number
of the shock is low, and injection of nuclei may be suppressed. This
case is considered in more detail in section \ref{sec:injection}.

The more realistic possibility for suppression of nuclei is given by
the case of ultra-relativistic shocks with large Lorentz factor (see
section \ref{sec:injection} for a detailed discussion). The fraction
of heavy nuclei $\lambda$ can be suppressed exponentially in this
case, see Eq.~(\ref{A/p}).

Finally, UHE nuclei can be photo-dissociated
in the photon field of a source \cite{Sigl,Berez05}.

\section{Transition from galactic to extragalactic cosmic rays}
\label{transition}

From the analysis of the dip we obtained the indications that the
transition from galactic to extragalactic cosmic rays takes place
at energy $E_{\rm cr} =1\times 10^{18}$~eV. The first indication
is given by the measured modification factor which at this
energy becomes larger than one (see Fig.~\ref{fig:dips}) in
contradiction with the definition of $\eta$ that forces $\eta (E) \leq
1$. This signals the appearance of a new component at lower energies,
which can be nothing but galactic cosmic rays.

The second indication is given by the low-energy 'diffusion cutoff'
of the extragalactic spectrum (Fig.~\ref{fig:diffuse}), which
inevitably provides the dominance of the galactic component at energy
$E < E_{\rm cr}$.

The energy at which the transition begins, $E_{\rm cr}$, is
completely determined by the equality of the rate of pair production
losses and the rate of adiabatic losses at the energy $E_{\rm eq}=
2.3\times 10^{18}$ eV \cite{BGG}. As stressed in
Section~\ref{discrete} the connection between $E_{\rm cr}$  and
$E_{\rm eq}$ is given by $E_{\rm cr} =E_{\rm eq}/(1+z_{\rm eff})^2$,
and results in $E_{\rm cr} \approx 1\times 10^{18}$~eV. The
transition must be observationally visible at some lower energy. It
coincides thus with the second knee located at energy $E_{\rm 2kn}$,
for which the  different experiments give $E_{\rm 2kn} \sim (0.4 -
0.8)\times 10^{18}$ eV. For the other models, where the transition
also occurs at the second knee, see \cite{Biermann}-\cite{Dermer}.

The explanation of the transition given above can be put in a simple
and general way: from the plot of the modification factors
in Fig. \ref{fig:dips}, one can see that the spectrum of
extragalactic cosmic rays reproduces the generation spectrum when
the modification factor tends to unity. The generation spectrum
is always flatter
than the spectrum of galactic cosmic rays $\propto E^{-3.1}$. This
argument is further strengthened by the low-energy  'diffusion cutoff'
in the extragalatic spectrum.

\begin{figure}[ht]
\begin{minipage}[h]{9cm}
\centering
\includegraphics[width=86mm,height=70mm,clip]{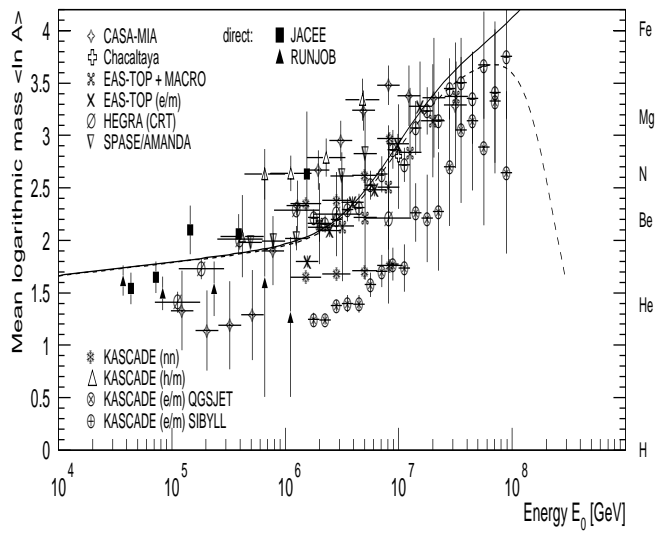}
\end{minipage}
\caption{Mean logarithmic mass number of cosmic rays as a function of
  energy \cite{horandel}.}
\label{fig:masses}
\end{figure}

\begin{figure}[ht]
  \begin{center}
    \includegraphics[width=14.0cm]{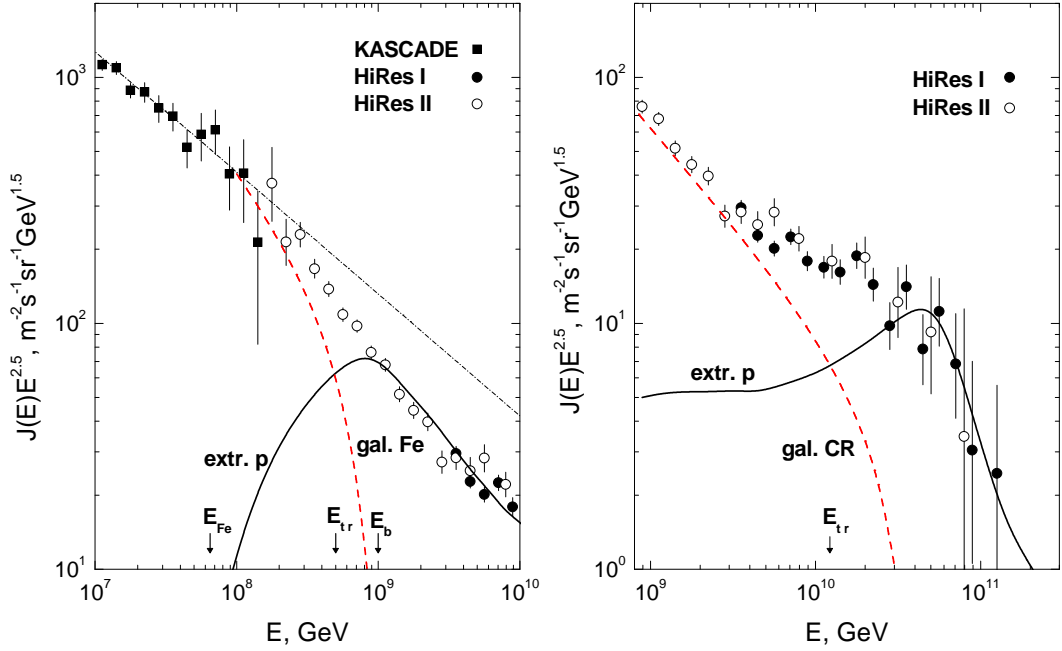}
\end{center}
  \caption{{\it Left panel: the second-knee transition }. The
    extragalactic proton spectrum is shown for $E^{-2.7}$ generation
    spectrum and for propagation in magnetic field with
    $B_c= 1$~nG and $l_c=1$~Mpc, with the Bohm diffusion at $E \lsim E_c$.
    The distance between sources is $d=50$~Mpc. $E_b=E_{\rm cr}=1\times
    10^{18}$~eV is the beginning of the transition, $E_{\rm Fe}$ is
    the position of the iron knee and $E_{\rm tr}$ is the energy where
    the galactic and
    extragalactic fluxes are equal. The dash-dot line shows the
    power-law extrapolation of the KASCADE spectrum to higher
    energies, which in fact has no physical meaning, because of the
    steepening of the galactic spectrum at $E_{\rm Fe}$.
    {\it Right Panel: the ankle transition},
    for the injection spectrum of extragalactic protons $E^{-2}$.
    In both cases the dashed line is obtained as a result of
    subtracting the extragalactic spectrum from the observed
    all-particle spectrum.}
  \label{fig:transition}
\end{figure}

The {\em dip-transition} model works most naturally in the framework
of the rigidity
model for the origin of the first knee. There are two versions of
this model: rigidity-confinement model and rigidity-acceleration model
(e.g. \cite{Biermann}). In the first model, the position of the proton
knee ($E_p \approx 2.5\times 10^{15}$~eV) is determined by magnetic
confinement of protons in the galactic halo. In this case the knee in
the spectrum of nuclei with charge $Z$ is located at $E_Z=ZE_p$.
The highest energy knee, due to iron, must be located at
$E_{\rm Fe}= 6.5\times 10^{16}$~eV. This finding seems to be confirmed
by Fig.~\ref{fig:masses}, where the mean $A$ is that corresponding to
iron nuclei at roughly this energy. At energy $E > E_{\rm Fe}$ the
total galactic flux, consisting mostly of iron, must be steeper. The
acceleration rigidity model predicts the same behaviour of the
spectra, but the knees appear due to different maximum acceleration
energies $E_{\rm max}$ for different nuclei. The data of KASCADE
confirm well the rigidity models when the SYBILL interaction model
is adopted, less well with the QGSJET model \cite{kascade}.
In the framework of a rigidity dependent origin of the knees, the
dip model describes the transition as the intersection of a steep
galactic spectrum at $E> E_{\rm Fe}= 6.5\times 10^{16}$~eV  with a
flat extragalactic proton spectrum at $E < E_{\rm cr}=1\times
10^{18}$~eV. Numerically this transition is shown in the left panel
of Fig.~\ref{fig:transition} for the Bohm diffusion of extragalactic
protons at energies below $E_{\rm cr}=1\times 10^{18}$~eV.

The beginning of the galactic steepening at $E_{\rm Fe}= 6.5\times
10^{16}$~eV and the extragalactic flattening below $E_{\rm
cr}=1\times 10^{18}$~eV are the results of separate pieces of
Physics and nothing "unnatural" takes place in the proximity of
these energies (they differ by one order of magnitude). The ratio of
the fluxes $J_{\rm gal}(E_{\rm Fe})/J_{\rm extr}(E_{\rm cr}) \sim
7\times 10^{3}$ is large, and we do not see any fine-tuning in this
model of the transition.

The galactic and extragalactic fluxes become equal at $E_{\rm
tr}=5\times 10^{17}$~eV. The transition is very prominent, if the iron
and proton components are resolved, but in the total spectrum the
transition appears as a faint feature known as the {\em second
knee}. This property is the same as for the other knees (proton,
helium, carbon, etc.) observed by KASCADE: while the transition
between the knees is distinct in the spectra with fixed chemical
composition, the resultant total spectrum has a smooth power-law
shape.

For the fraction of
iron and proton fluxes in the energy range $1\times 10^{17} -
1\times 10^{18}$~eV see \cite{BGH,AB1}.

The right panel of Fig.~\ref{fig:transition} shows the traditional
transition from galactic to extragalactic cosmic rays at the ankle
($E_a \approx 1\times 10^{19}$~eV). In this model the extragalactic
component has a very flat generation spectrum $\propto E^{-2}$ which
naturally intersects the steep ($\propto E^{-3.1}$) galactic
component. The most attractive feature of this model is given by the
flatness of the extragalactic generation spectrum, which provides
reasonable luminosities of the sources and a natural interpretation
of the intersection of the galactic and extragalactic cosmic ray
components. Being stimulated by the discovery of the ankle by the
Haverah Park array in the '70s, this model has been considered
recently in Refs.~\cite{ankle,DeMSt}.  Both models of the transition,
at the second knee and at the ankle, have some advantages and
problems, as summarized below: 

\begin{itemize}

\item The second knee model is inspired by and based on the
numerical confirmation of the existence of the dip as a spectral
feature of extragalactic protons interacting with the CMB (see
Fig.~\ref{fig:dips}).  The probability of an accidental agreement,
estimated from the $\chi^2$, the number of free parameters and the
number of energy bins in each of the four experiments, is very
small. The ankle model explains the dip as a possible interplay
between galactic and extragalactic spectra. It looks rather odd that
such feature has exactly the same shape as that of the CMB-induced
dip.

\item  The explanation of the transition is more straightforward in
the ankle model: it is the simple intersection of the flat extragalactic
spectrum with the steep galactic spectrum. This model naturally
predicts a rather low luminosity of the sources and allows to
incorporate an arbitrary fraction of heavy nuclei in the total
flux at $E > 1\times 10^{19}$~eV, in case the future experiments
will show that this is needed. The second knee
transition is also based on the intersection of a steep galactic
spectrum with a flat extragalactic spectrum. The flatness of the
extragalactic spectrum (diffusion 'cutoff') appears quite naturally
at energy close to $E_{\rm cr} =1\times 10^{18}$~eV due to diffusion
of protons with $E < E_{\rm cr}$. However, a low luminosity of
the sources can be achieved only by postulating a distribution of
maximum energies at the sources. The energy where the {\it effective}
generation spectrum shows the steepening is a free parameter.

\item
The dip is modified by the presence of heavy nuclei in the primary
radiation and it allows only small admixture of heavy nuclei at the
dip and above it. This may in turn be interpreted as a possible
signature of the model of transition at the dip.

\item
The model of the transition at the ankle requires that the galactic
component of cosmic rays extends to energies in excess of $10^{19}$
eV. This requires a revision of the existing galactic models of
propagation and acceleration, which predict the maximum acceleration
energy to be less than $1\times 10^{18}$~eV for iron nuclei
\cite{Emax}. The model of transition at the dip appears to be in
agreement with observations of the first knee combined with a
rigidity dependent picture (either related to propagation or to
acceleration) of the knees for nuclei more massive than hydrogen.

\item
At energy $E \geq 1\times 10^{17}$~eV , i.e. much higher than the
proton knee, at least 10\% of the observed flux is in the form of
protons. This fraction is most naturally explained in the context of
the second knee model. It also represents a serious challenge for
the ankle model.

\end{itemize}

As stressed above, both models make clear predictions, that can and
should be used to prove or disprove them. The most critical
observation is the measurement of the chemical composition in the
energy region around $E \sim 1\times 10^{18}$~eV. While the dip
model predicts a strong dominance of protons, in the ankle model a
strong dominance of iron nuclei is expected. We think that the
fluorescence measurements provide us with the most promising tool
for discrimination of these two models. At present HiRes elongation
rate data \cite{mass-Hires} show the transition to a
proton-dominated mass composition at $E \approx 1\times 10^{18}$~eV
as the dip model predicts, while HiRes predecessor, Fly's Eye, shows
this transition at higher energies. Referring to this contradiction
we want to emphasize the uncertainties, both experimental and
theoretical, involved in present analysis, and which hopefully will
be overcome in the future.

As we discussed above, the ankle model in its canonical form
predicts a transition from galactic to extragalactic component at
energy $E_a \approx 1\times 10^{19}$~eV, as observational data from
Fig.~\ref{fig:dips} imply, and as most authors of the ankle models
assume. However, there could be an intermediate possibility between
the dip and the ankle models, when transition occurs at $1\times
10^{18} < E < 1\times 10^{19}$~eV. The most elaborated model of this
type is the mixed-composition model \cite{parizot}. The transition
is found to occur at  energy $E \sim 3\times 10^{18}$~eV, and thus
it softens the difficulties with the highest energy end of the
galactic cosmic ray spectrum. In this model, the observed dip, as 
it is shown in
Fig.~\ref{fig:dips}, is reproduced exactly, provided that the
galactic component of cosmic rays is fitted
{\it a posteriori} by subtracting the calculated extragalactic
component from the observed total spectrum (the model for the
galactic spectrum is not discussed and the reason why the dip in
this model coincides exactly with the dip calculated for
extragalactic protons is left unanswered). Inspired by
observations of galactic cosmic rays, the chemical composition of
extragalactic cosmic rays is assumed to be mixed. The authors show
that elongation rates, especially the ones from the data of Fly's
Eye and Yakutsk, confirm better the mixed model than the pure proton
model of the dip. We can add to this finding that Akeno data
\cite{mass-Akeno} confirm also the mixed composition, while data of
HiRes \cite{mass-Hires}, HiRes-MIA \cite{Hi-Mia} and more reliable
{\em muon data} of Yakutsk \cite{Glushkov00} support the
proton-dominated composition at $E \geq 1\times 10^{18}$~eV. Based
on these many contradictory sets of data we do not feel of sharing
with the authors of \cite{parizot} their trust in the accuracy of
interaction-dependent analysis of the elongation rate at present,
although we hope in its future progress.

\section{Conclusions}
\label{conclusions}

The dip is a feature in the extragalactic cosmic ray spectrum,
that originates from the Bethe-Heitler pair production of protons
on the cosmic microwave background  (\cite{BGG3},\cite{BGG}). The
dip appears at energy $1\times 10^{18} - 4\times 10^{19}$~ eV, with
shape practically independent of the discreteness and inhomogeneity in
the source distribution, independent of local overdensity and deficit
of the sources and of the maximum acceleration energy, independent
of the presence or absence of magnetic fields in the intergalactic space,
and independent of fluctuations in $p\gamma$ interactions.
The cosmological evolution of UHECR sources, with the parameters
 taken from observations of AGN, do not affect the dip. The only
phenomenon which modifies noticeably the shape of the dip is the presence
of a large fraction of nuclei, heavier than hydrogen, in the generation
spectrum. We consider the small fraction of nuclei, allowed by the dip
shape, as an indication of possible mechanisms of acceleration or
injection, operating in UHECR sources (Section \ref{sec:injection}).

The predicted shape of the dip is in excellent agreement with the
data of Akeno-AGASA, Fly's Eye, HiRes and Yakutsk detectors
(the data of Auger are inconclusive because of the absence of data at
$E< 3\times 10^{18}$~eV, essential for the dip). The energy
calibration of these detectors by the position of the dip results in
excellent agreement between the measured fluxes.

To our knowledge, the precision of the agreement between the predictions
for the dip, which need only two free parameters, and observations
(see Fig.~\ref{fig:dips}) is the best that ever existed in cosmic ray
physics.  This, and the results of energy calibration of the detectors
by the position of the dip (see Fig.~\ref{fig:AgHiYa}), makes
improbable that we are observing an accidental agreement between
the predictions and the observations.

All the discussion presented so far in the paper was based on purely
phenomenological and largely model independent grounds. Clearly, in
order to establish a connection with existing theories of
acceleration or propagation of cosmic rays, the dip needs to be
related to models. In passing we notice however that the shape of
the dip agrees well with data when AGN with their observed
cosmological evolution are considered as sources of UHECR (Section
\ref{evolution}). A possible, though model dependent, correlation
between the maximum acceleration energy and the luminosity of the
sources might also solve the problem of the excessive energetics
required by the dip scenario (Section \ref{Emax}). A small fraction
of heavy nuclei can also be accommodated in the model (Section
\ref{chemical}).

In the dip scenario the transition from galactic to extragalactic
cosmic rays takes place at $E_{\rm cr} \approx 1\times 10^{18}$~eV.
The natural character of this transition is guaranteed by the fact
that a flat extragalactic spectrum at $E < E_{\rm cr}$ intersects a
steep galactic spectrum at $E > E_{\rm Fe} \sim 1\times 10^{17}$~eV.
Observationally the transition occurs at the second knee.

An alternative possibility for the transition is given by the {\em
ankle} model (see Section \ref{transition}).

The ankle model has many attractive features. It gives a simple and
natural picture of the transition as the intersection of a steep galactic
spectrum with a very flat extragalactic spectrum, resulting from a
generation spectrum as predicted by the 'standard' acceleration
models ($E^{-2}$ for non-relativistic shock acceleration and
$E^{-2.2}$ for relativistic shocks). These generation spectra
have no problem with energetics. The model can also easily accommodate
a large fraction of heavy nuclei, if observations show that this is
needed.

On the other hand the ankle model has weaknesses: the model requires
that the galactic component extends to energies in excess of $10^{19}$
eV, in apparent contradiction with the KASCADE data. We add also
that it seems to contradict the most current models of cosmic
ray acceleration in galactic sources \cite{Emax}. The model also has
difficulties in explaining the $\sim 10\%$ of protons observed
in Akeno at $E \sim 10^{17}$~eV (in the dip model these protons are
extragalactic).

In conclusion, we want to emphasize the importance of having
detectors working in the energy range $3\times 10^{17} - 1\times
10^{19}$~eV, such as possible low-energy extensions of Auger,
Telescope Array (TA) and low-energy extension of TA (TALE). These
detectors can reliably solve the problem of measuring the mass
composition in the transition region. It is important to observe
that the two basic models of the transition, at the second knee and
at the dip, give drastically different predictions for the chemical
composition in the energy interval  $1\times 10^{18} - 1\times
10^{19}$~eV: proton-dominated in the dip model and iron-dominated in
the ankle model. The updated fluorescent method with measurement of
atmosphere transparency for each event can reliably distinguish
these two models, in spite of existing experimental and theoretical
(models of interaction) uncertainties. 

One can add another signature common to both the dip and the ankle
scenarios (see \cite{parizot}) namely the appearance of a jump in $x_{max}$ at
energy $E_{tr}$, which equals $\sim 5\times 10^{17}$ eV for the dip
model and $\sim 10^{19}$ eV for the ankle model. This feature appears
because of a sharp transition from a steep galactic (iron dominated)
spectrum to a flat extragalactic (proton dominated) component (see
Fig. 10). Such a jump is practically absent or much smoother in the
mixed composition model. In contrast with the dip and ankle models,
the mixed composition model needs a
higher accuracy of the fluorescent method to distinguish the mixed
extragalactic composition from the proton-dominated or
iron-dominated compositions (there are no reliable predictions for
the mass composition of extragalactic UHECR).

One should keep in mind the great predictive power of the {\em
spectrum measurements} for the determination of the mass composition
of extragalactic UHECR. The modification factors for protons and
different nuclei (see Fig.~\ref{fig:dip-nucl} for comparison) differ
very strongly in the energy region $1\times 10^{18} - 1\times
10^{20}$~eV and even small admixtures of nuclei can be discovered as
a distortion of the proton modification factor. Note that this
possibility exists for all models: dip, ankle and intermediate
transition.

In principle, the {\em anisotropy} discriminates the different
models of transition. While the dip-transition model predicts an
extragalactic anisotropy at $1\times 10^{18} - 1\times 10^{19}$~eV,
the ankle models predicts the galactic anisotropy at this energy
range. However, in the case of an expected iron-dominated
composition of galactic cosmic rays both models predict too small
anisotropy.

We think that measurements of the spectrum and chemical composition
in the energy region $3\times 10^{17} - 1\times 10^{19}$~eV can
resolve the existing problem of establishing where the transition
from galactic to extragalactic cosmic rays takes place and much
experimental effort should be put in aiming at this goal.
\section*{Acknowledgments}
We thank ILIAS-TARI for access to the LNGS research infrastructure and for
the financial support through EU contract RII3-CT-2004-506222. The
work of S.G. is partly supported by Grant No. LSS-5573.2006.2.
\appendix
\section{Theoretical issues: the dip scenario and the acceleration of
  extragalactic cosmic rays}
\label{theory}

One of the arguments that is often put forward as being in favor of
the ankle scenario for the transition from galactic to extragalactic
cosmic rays is that the ankle scenario requires injection spectrum
$E^{-2}-E^{-2.3}$, compatible with the prediction from the theory of
particle acceleration at shocks. In this Section we discuss
the acceleration aspects for the dip scenario, demonstrating the existence
of quite reasonable possibilities. Our discussion will concern mainly the
shock acceleration mechanism.

\subsection{The spectrum of particles accelerated at non relativistic
  shocks}
\label{nonrel}

The {\it test particle} theory of particle acceleration at
collisionless shocks makes very clear predictions for the spectrum of
accelerated particles. For a shock moving with Mach number $M$, the spectrum
is a power law at all momenta larger than the injection momentum, and
the slope is related to the Mach number as $\gamma=2(M^2+1)/(M^2-1)$
for adiabatic index 5/3.
When the shock becomes transonic, the efficiency of particle
acceleration vanishes and the spectrum becomes infinitely steep. If
the shock moves at very high (but still non relativistic) speed, the
spectrum tends asymptotically to $E^{-2}$, for $M\to \infty$. This is
the case of {\it strong shock} which is commonly used in the
literature. As long as the shock is non-relativistic, these
results remain true irrespective of the choice of the diffusion
properties in the shocked fluid, which instead are important for the
determination of the maximum energy of the accelerated particles.
It is worth stressing that a slope $2.7$ would correspond in this
context to a Mach number $M=2.6$. An instance of a situation in which
weak shocks like these develop are mergers of clusters of galaxies
\cite{gabici}, though the maximum energy in this case is not
expected to exceed $10^{19}$ eV \cite{blasi2001}.

The predicted spectrum of particles accelerated at non relativistic
shocks is substantially changed when the approximation of {\it test
  particles} is relaxed \cite{malkov1,malkov2,blasi1,blasi2}: the
effect of the dynamical reaction of particles onto the shock is to
steepen the low energy part of the spectrum and flatten the highest
energy part compared with the {\it test particle} prediction.
For strongly modified shocks the slope of the spectrum at the highest
energies can become as flat as $E^{-1.2}$ \cite{amato}.

\subsection{Acceleration and Reacceleration}
\label{reacc}

The spectrum of cosmic rays accelerated in a region which is crossed
by several shock waves of different strength (e.g. different Mach numbers)
may be quite complex. The case
of two shocks is illuminating in this respect. Let us consider the
case of a shock wave with Mach number $M_1$, which accelerates
particles to a power law spectrum with slope $\gamma_1$ up to a
maximum energy $E_1$. If a second shock front with Mach number
$M_2$ passes through the same region, the pre-existing population
of cosmic rays is re-energized. It can be easily shown \cite{bell78}
that the final spectrum of cosmic rays at large momenta is a new power
law with slope close to $\gamma_1$ if $M_2<M_1$, or approaching
$\gamma_2$ if $M_2>M_1$. If however the second shock front is able to
achieve a maximum energy $E_2>E_1$ and $M_2<M_1$, then the final
spectrum is $E^{-\gamma_1}$ for $E<E_1$ and $E^{-\gamma_2}$ (steeper
than at low energy) for $E_1<E<E_2$. For instance if $M_1=100$ and
$M_2=2.6$, the final spectrum will be a broken power law with low
energy slope $\sim 2$ and high energy slope $\sim 2.7$. This exercise,
of didactic interest, shows that the assumption of a single power law
spectra, usually adopted for simplicity may be far from realistic situations.

\subsection{The spectrum of particles accelerated at ultra-relativistic
  shocks}
\label{rel}

The spectrum of the particles accelerated at shocks moving with
relativistic speed is not universal, in that it depends on the details
of the scattering properties of the plasma in which the shock
develops \cite{vietri,blasivietri,morlino1} and the equation of state of the
downstream plasma \cite{morlino2}. Some sort of universality is
achieved if the motion of the particles occurs in the regime of small
pitch angle scattering and if the shock has Lorentz factor
$\Gamma_{sh}\gg 1$ (see \cite{blasivietri} and references therein). In
this special case, the spectrum of the accelerated particles is a power
law with slope $2.32$ \cite{gallant} with a low energy cutoff at
energy $\sim \Gamma_{sh}^2 m_p c^2$, if $m_p$ is the mass of the
accelerated particles.

Many physical phenomena affect this simple prediction, as discussed in
detail in \cite{blasivietri}. The spectrum has a slope $\sim 2.7$ if
the relativistic equation of state for the downstream gas is assumed
and the shock speed in units of the speed of light is $0.9$
\cite{blasivietri}. Spectra flatter than $2.32$ can easily be
obtained when the scattering of the particles is assumed to be in the
regime of large angle scattering. Recently the authors of \cite{lemsh}
noticed that an isotropic Kolmogorov turbulence in the upstream
plasma, compressed by the shock front in the direction parallel to the
shock surface leads to a spectrum of accelerated particles with slope
$2.7$. The effect is due to the fact that the scattering in the
downstream plasma becomes anisotropic. The general case of anisotropic
scattering was recently studied in detail in \cite{morlino1}. All these
cases show that the spectrum of the accelerated particles at a
relativistic shock front is determined by many poorly known aspects of
the microphysics of the scattering, and that universality can only be
used as a sort of {\it rule of thumb}.

\subsection{Other acceleration mechanisms}

Many processes lead to particle acceleration, with spectra and
efficiencies that depend on the specific situation. We briefly mention
here two cases: particle acceleration driven by a relativistic wind,
and particle acceleration through pinch instability.

The first acceleration mechanism was discussed in the context of the
generation of UHECR in \cite{epstein} and \cite{arons} and applied to
the case of magnetized rapidly spinning neutron stars. Charged
particles trapped on the open magnetic field lines on the star (namely
outside the light cylinder) are pushed outwards with Lorentz factor
close to that of the relativistic wind induced by the magnetic
pressure. In this scenario, a spectrum of particles is generated due
to the slowing down of the spinning neutron star. If the slowing down
is dominated by magnetic dipole radiation, it was shown in
\cite{epstein} that the spectrum of the accelerated particles is $\propto
E^{-1}$, therefore suitable only to explain the AGASA excess, if it is
there.

Particle acceleration in jets due to the pinch mechanism was suggested
first for tokamaks (where it was confirmed on the laboratory scale)
and was then rescaled to astrophysical objects in \cite{pinch}.
The pinch mechanism of acceleration works due to the pinch-neck
instability, which is accompanied by the generation of intense
electric fields. The solution of the kinetic equation \cite{pinch}
predicts a power-law spectrum of accelerated particles
$Q(E) \propto E^{-\gamma}$~ with~ $\gamma =1 + \sqrt{3}$.

The maximum energy at the source may exceed $10^{20}$ eV, if the
results of laboratory experiments are extended to jets typical of
AGNs.

\subsection{A model for complex injection}
\label{Emax}

In this section we discuss the possibility, first put forward in
\cite{michael}, that the sources of UHECR might be characterized
by parameters of the spectrum of accelerated particles that differ
from source to source. In particular,
the maximum energy might depend on environmental parameters of the
sources, and not only on the type of sources. The implication of this
assumption is that the {\it effective} injection spectrum
due to the sum over all the sources at
the given redshift may appear as having different slopes in different
energy ranges.
\begin{figure}[ht]
\begin{minipage}[h]{10cm}
\centering
\includegraphics[width=86mm,height=70mm,clip]{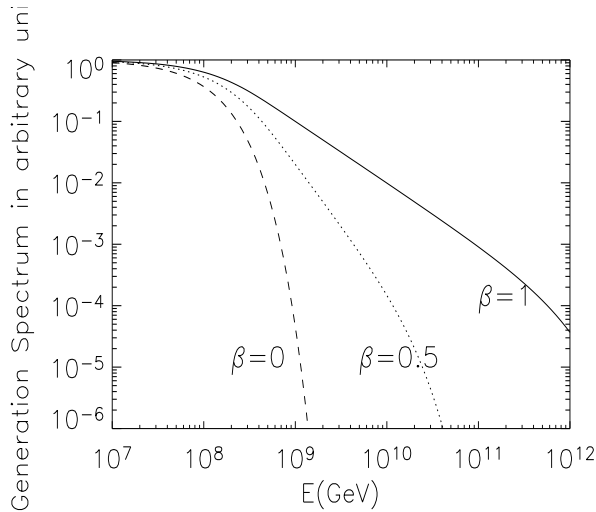}
\end{minipage}
\caption{\label{fig:emax} Normalized injection spectrum
$Q(E)(E/E_0)^\gamma$ for $\beta=0,~0.5,~1$.}
\end{figure}
A simple example may serve to clarify the basic idea: let us assume
that a class of sources of UHECR has a luminosity function given by
$$N(L)=N_0 \left(\frac{L}{L_{min}}\right)^{-\alpha},$$
and that the maximum energy of the sources correlates with the source
luminosity as
$$E_{max}(L)=E_{max}^{*}\left(\frac{L}{L_{min}}\right)^{\beta},$$
where $E_{max}^{*}$ is the maximum energy of the sources with luminosity
$L_{min}$ (in the general case the function $N(L)$ might depend
explicitly on redshift). Moreover, let us assume that the injection
spectrum of each source has the form of a power law with an
exponential cutoff:
$$Q_s(E) = \frac{(\gamma_g-2)L}{E_0^2}
\left(\frac{E}{E_0}\right)^{-\gamma_g}
\exp\left[-\left(\frac{E}{E_{max}(L)}\right)\right].$$
The {\it effective} injection spectrum,
i.e. the number of particles generated per unit comoving volume
at the redshift $z$ is therefore
\beq
Q(E) = \int_{L_{min}}^{L_{max}} ~ dL ~ N(L) Q_s(E) =
Q_0 \left(\frac{E}{E_0}\right)^{-\gamma_g}
\int_{1}^{r} dx ~ x^{-\alpha+1}
\exp\left[-\left(\frac{E}{E^{*}_{max} x^{\beta}}\right)\right],
\eeq
where $r=L_{\rm max}/L_{\rm min}$ and
$Q_0=(\gamma_g-2)N_0(L_{\rm min}/E_0)^2$.

For illustrative purposes we choose $E_{max}^{*}=10^{17}$ eV, $r=10^4$
and $\alpha=3$. In Fig. \ref{fig:emax} we plot $Q(E)(E/E_0)^\gamma$ in
arbitrary units for $\beta=0,~0.5,~1$.

One can see that despite the fact that each source has a power law
injection spectrum with given slope $\gamma$, the effective injection
relevant for the calculations of the propagation of UHECR has the
shape of a broken power law, with steeper slope at high energies.
\begin{figure}[ht]
  \begin{center}
    \includegraphics[width=8.0cm]{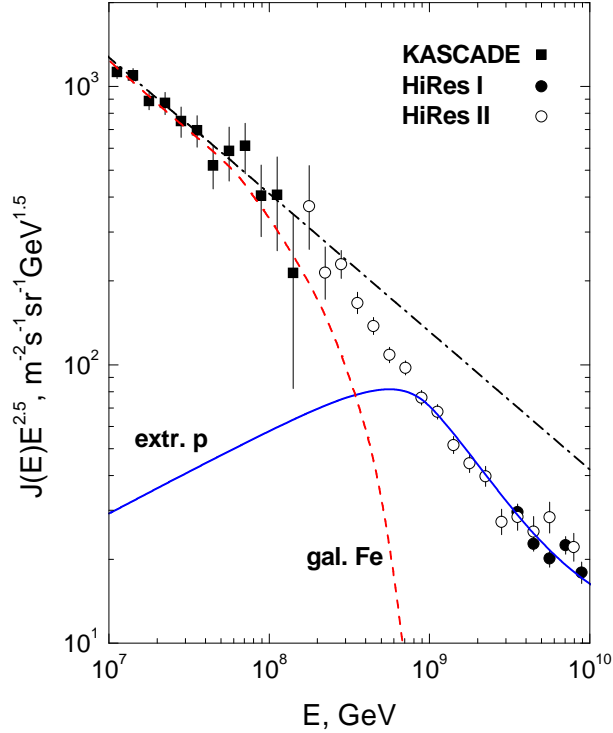}
\end{center}
  \caption{Spectrum of the extragalactic cosmic rays obtained with the
  complex injection spectrum having slope $\gamma=2.2$ for $E<E_c$ and
  slope $\gamma=2.7$ for $E>E_c$ ($E_c=10^{18}$ eV). The dashed line
  is obtained as a result of subtracting the extragalactic spectrum
  from the observed all-particle spectrum.}
  \label{fig:transitionEc}
\end{figure}
The attractiveness of this model is that it allows us to adopt a
generation spectrum for each source that is a power law with slope
$2-2.4$ and obtain the {\it desirable} injection spectrum at high
energy due to the superposition of many sources. The flat spectrum
alleviates or solves the problem with energetics discussed in
Sec. \ref{dip} (see subsection \ref{energetics}). On the other hand
it becomes very difficult to
make realistic predictions in this model, unless the sources of UHECR,
the acceleration process and the evolution properties of the sources
are very well known.

In Fig. \ref{fig:transitionEc} we plot the predicted spectrum of
extragalactic cosmic rays in the case of an injection spectrum which
is $E^{-2.7}$ for $E>E_c=10^{18}$ eV and $E^{-2.2}$ for $E<E_c$. The
dashed line is the spectrum of galactic cosmic rays inferred by
subtraction of the extragalactic flux from the all-particle observed
spectrum. Again the transition takes place at energy $E_{\rm tr}
\approx 3.5\times 10^{17}$~eV.

\subsection{Injection to shock acceleration and nuclei-to-protons ratio
in the accelerated flux}
\label{sec:injection}

The agreement of the dip with observations for the best fit parameters
implies the strong dominance of protons in the primary radiation. Meanwhile,
the observed abundance of heavy nuclei in galactic cosmic rays is
large enough to affect the good agreement of the proton dip with the
data. Since the sources of UHECR are different from those for galactic
cosmic rays, one may ask: Can acceleration mechanisms which operate in UHECR
sources result in lower abundance of heavy nuclei?

Any source acceleration mechanism operates in the gas where heavy
nuclei are present. The largest and guaranteed abundance is given by
cosmological helium which results in the ratio of densities
$n_{\rm He}/n_{\rm H}= 0.079$. We will focus on the He/p
ratio in the acceleration flux, limiting our discussion
to shock acceleration.

The He/p ratio in cosmic rays differ from the density ratio
$n_{\rm He}/n_{\rm H}= 0.079$ due to the injection process.

The injection of charged particles in collisionless
shocks is a complex and still unsolved plasma physics problem.
The shock is most likely formed in a collisionless plasma due
to Weibel instability and its thickness is reasonably thought to be of
the order of a few Larmor radii of the thermalized particles in the
downstream plasma. This is however a quite heuristic argument based on
the fact that Weibel instability generates wild magnetic fluctuations
in the shock structure (even in the absence of an initial magnetic
field). If these fluctuations are such that their amplitude exceeds
the background field (this is certainly the case if there is no
initial magnetic field), the thickness of the shock can
be smaller than suggested by our heuristic argument. We are not aware
of accurate predictions of the internal structure of a collisionless
shock in the literature. This makes the investigation
of the injection process very uncertain at the present time.

The energetics of the shock is expected to be dominated by protons,
which are heated downstream to a temperature $T$
which roughly corresponds to the isotropization of the momenta of
upstream particles flowing in the shock rest frame as the cold fluid.
The average momentum of the thermal protons is therefore $p_{th}^{(p)}
\approx m_p v_s \gamma_{sh}$, where $v_s$ and $\gamma_{sh}$ are the
velocity and Lorentz factor of the upstream fluid in the shock frame.
For a non relativistic shock this reduces to $p_{th}^{(p)}\approx
m_p v_s$, which is approximate since it neglects the bulk
motion of the downstream plasma in the shock frame.
According with the argument given above, the thickness of the
shock is $R_{sh}\approx \xi p_{th}^{(p)}c/(eB)$, with $\xi=2-4$
\cite{vannoni}.

We consider now nuclei with mass number $A$ and charge $Z$ in the
downstream region. In order to cross the shock front and proceed
towards upstream, their Larmor radius $R_L=pc/ZeB$
must exceed $R_{\rm sh}$, and hence only nuclei with momenta higher than
$p_{\rm inj}^A=ZeBR_{\rm sh}/c$  can be accelerated. In this
way we arrive at the rigidity-dependent injection mechanism,
when injection momenta for nuclei and protons are related by
\beq
p_{\rm inj}^A = Z p_{\rm inj}^p .
\label{inj}
\eeq
This simple argument suggests that the injection of nuclei
is easier than the injection of protons, in agreement with what
is usually assumed in the current literature \cite{ellison}.
It is however difficult to quantify this difference on theoretical
grounds. Based on the observations of cosmic rays in the Galaxy,
nuclei are systematically more abundant in cosmic rays than they
are in the interstellar medium in the solar neighborhood
\cite{ellison,berez}.

If we assume, as suggested by Eq.~(\ref{inj}) that $p_{\rm inj}^A = Z
p_{\rm inj}^p$, and we also assume that the injection spectrum is a power
law in momentum up to a maximum momentum, then the
number of accelerated nuclei with momentum $p$ is
$Q_A(p)=K_A(p/p_{\rm inj}^A)^{-\gamma_g}$, where $A=1$ corresponds to
protons, $K_A=\epsilon_A n_A(\gamma_g-1)/p_{\rm inj}^A$ with
$\epsilon_A$ being the fraction of thermal nuclei involved in
acceleration. For large enough momenta $p$ one obtains
\beq
\frac{Q_A(p)}{Q_p(p)} = \frac{\epsilon_A}{\epsilon_p}
\frac{n_A}{n_H}Z^{\gamma_g-1},
\label{Q_A/Q_p}
\eeq
where $\epsilon=\epsilon_A/\epsilon_p\geq 1$ depends on the distribution
of thermal nuclei at momenta $\geq p$.

From Eq.~(\ref{Q_A/Q_p}) it  follows that in most cases the presence
of He nuclei (Z=2) with cosmological number density $n_H$ results
in a too strong distortion of the proton modification factor. An
exception may be given by the case of acceleration in a cold
gas. We remind the reader that the first and second
potentials for helium are very high, 24.6 eV and 54.4 eV,
respectively. In the extreme case of neutral helium the mixed
modification factor will be $\eta(E) \approx \eta_p(E)$. If the
temperature of the gas is low enough to prevent a second ionization,
the condition for injection is given by $Z_{\rm eff}=1$ and the
dangerous term $Z^{\gamma_g-1}$ in Eq.~(\ref{Q_A/Q_p}) disappears.
Note, that at larger energies, the He nuclei become fully ionized,
but the total number of accelerated nuclei is determined by the
injection condition with $Z_{\rm eff}=1$.

Let us illustrate this possibility by using the case of a
non-relativistic shock with Mach number $M$ propagating in a warm
upstream gas with temperature $T_u$. Injection occurs in the
downstream region, where the temperature $T_d$ of the gas found from
the Rankine-Hugoniot relations is given by \beq
\frac{T_d}{T_u}=\frac{1}{16} (5M^2-1)(1+\frac{3}{M^2}), \label{temp}
\eeq valid for a gas with adiabatic index 5/3. In the limit of large
Mach number $M \gg 1$~~ $T_d/T_u = (5/16)M^2$ and the spectral index
of accelerated particles tends to $\gamma_g=2$. In the case of a
small Mach number $\gamma_g=2(M^2+1)/(M^2-1)$ and the increase of
$T_d$ is modest. For example, in the case of $M=2.59$,~
$\gamma_g=2.7$, as needed to fit the dip, and the ratio of
temperatures $T_d/T_u=2.9$. In the case $T_u \sim 10^{4} - 10^5$~K
the downstream temperature is below the second ionization potential
and injection of He-nuclei is suppressed. For the case of $M=4$ when
compression $\sigma$ is close to 4:  $\sigma=3.4$,~ $\gamma_g=2.3$ and
$T_d/T_u=5.9$.

Another example when the fraction of accelerated nuclei might be suppressed
is given by relativistic shock acceleration with large $\Gamma_{\rm sh}$.
In this case the cold upstream particles have a downstream energy
$E_d \sim \Gamma_{\rm sh} A m_p$ and the temperature of the downstream
gas $T$ can be very high. We shall assume that the temperature of nuclei
and protons is the same $T_A \approx T_p$. This assumption is based
on poorly known physics of thermalization of the gas downstream. The
thermalization occurs most probably due to collective processes
and it is not known how fast these processes are. If they are slow,
the equilibrium condition  $T_A \approx T_p$ can be reached on average
too far from the shock front for efficient injection to take place.
Our assumption implies a fast thermalization. Thus we assume that
particles downstream have a thermal distribution
\beq
n_A^{\rm th}(p)=\frac{n_A}{2T^3}p^2 \exp(-p/T) ,
\label{th}
\eeq
where $n_A$ is the momentum integrated density and A=1 corresponds to
protons. Particles are injected in the regime of acceleration if
$p\geq p_{\rm inj}^A$. The density of accelerated particles is
$n_A^{\rm acc}(p)=K p^{-\gamma_g}$ at $p\geq p_{\rm inj}^A$ and the
condition of particle number conservation reads
$n_A^{\rm th}(\geq p_{\rm inj}^A)= n_A^{\rm acc}(\geq p_{\rm inj}^A)$.
Solving these equations one obtains
\beq
\frac{n_A^{\rm acc}(p)}{n_p^{\rm acc}(p)}=
\frac{n_A}{n_H}Z^{\gamma_g+1}\exp \left [ -\frac{p_{\rm inj}^p}{T}(Z-1)
\right ]
\label{A/p}
\eeq
For $p_p^{\rm inj} > 2 T$ the fraction of nuclei with $Z \geq 2$ is
exponentially suppressed.


\begin{thebibliography}{99}

\bibitem{GZK}
K. Greisen, Phys.\ Rev.\ Lett. {\bf 16}, 748 (1966);\\
G. T. Zatsepin, V.A. Kuzmin, Pisma Zh.\ Experim.\ Theor.\ Phys.\ {\bf 4},
114 (1966).

\bibitem{HS85}
C. T. Hill and D. N. Schramm, Phys.\ Rev.\ D {\bf 31}, 564 (1985).

\bibitem{Stanev00}
T. Stanev \etal, Phys.\ Rev.\ D {\bf 62}, 093005 (2000).

\bibitem{BG88}
V. S. Berezinsky and S. I. Grigorieva, Astron.\ Astroph.\ {\bf 199}, 1
(1988).

\bibitem{BGG3}
V. Berezinsky, A. Z. Gazizov, S. I. Grigorieva,
Phys. Lett. B {\bf 612}, 147 (2005) [astro-ph/0502550].

\bibitem{BGG}
V. Berezinsky, A. Z. Gazizov, S. I. Grigorieva, hep-ph/0204357.

\bibitem{agasa}
K. Shinozaki and M. Teshima, Nucl. Phys. B (Proc. Suppl.) {\bf 136},
18 (2004).

\bibitem{FE}
D. J. Bird \etal, [Fly's Eye collaboration], Ap.\ J. {\bf 424},
491 (1994).

\bibitem{hires}
HiRes collaboration,  Phys. Rev. Lett. {\bf 92}, 151101 (2004).

\bibitem{Yakutsk}
A. V. Glushkov \etal \ [Yakutsk collaboration]
Proc. of 28th Int. Cosmic Ray Conf. (Tsukuba, Japan), {\bf 1}, 389 (2003);

\bibitem{NW}
M. Nagano and A. Watson, Rev.\ Mod.\ Phys.\ {\bf 72}, 689 (2000).

\bibitem{ankle}
A. M. Hillas, preprint astro-ph/0607109;
A. M. Hillas, J.Phys. G: Nucl. Part, Phys. {\bf 31}, R95 (2005);
A. M. Hillas, Nucl. Phys. B (Proc. Suppl) {\bf 136}, 139 (2004);
E. Waxmam, Phys. Rev. Lett. {\bf 75}, 386 (1995);
M. Vietri, Astroph. J. {\bf 453}, 883 (1995);
T. Wibig, A. W. Wolfendale, J. Phys. G {\bf 31}, 255 (2005).

\bibitem{DeMSt}
D. De Marco and T. Stanev, Phys. Rev. D {\bf 72}, 081301 (2005).

\bibitem{AB}
R. Aloisio and V. Berezinsky, Astroph. J. {\bf 612}, 900 (2004).

\bibitem{Lem}
M. Lemoine, Phys. Rev. D {\bf 71}, 083007 (2005).

\bibitem{AB1}
R. Aloisio and V. Berezinsky, Astroph. J. {\bf 625}, 249 (2005).

\bibitem{Biermann}
P. Biermann et al., astro-ph/0302201 (2003).

\bibitem{Hoerandel}
J. R. Hoerandel, Astropart.\ Phys.\ {\bf 19}, 193 (2003).

\bibitem{Dermer}
S. D. Wick, C. D. Dermer, A. Atoyan, Astropart.\ Phys.\ {\bf 21}, 125 (2004).

\bibitem{mass-Hires}
R. U. Abbasi \etal [HiRes collaboration] Astroph. J. {\bf 622}, 910 (2005).

\bibitem{Hi-Mia}
T.Abu-Zayyad et al.,Phys. Rev. Lett. {\bf 84}, 4276 (2003).

\bibitem{Glushkov00}
A. V. Glushkov {\it et al.} (Yakutsk collaboration)\ JETP Lett.\
{\bf 71}, 97 (2000).

\bibitem{HP}
M. Ave \etal [Haverah Park collaboration], Astropart.\ Phys.\ {\bf 19},
61 (2003).

\bibitem{mass-Akeno}
M. Honda et al [Akeno collaboration], Phys. Rev. Lett. {\bf 70}, 525 (1993).

\bibitem{Watson04}
A. Watson, Nucl. Phys. B (Proc. Suppl.) {\bf 136}, 290 (2004).

\bibitem{ssc}
M.~Takeda [AGASA collaboration] , Astrophys. J. {\bf 522}, 225 (1999).

\bibitem{ssa1}
P. Blasi, D. De Marco, Astropart. Phys. {\bf 20}, 559 (2004);

\bibitem{ssa2}
M. Kachelriess, D. Semikoz, Astropart. Phys. {\bf 23}, 486 (2005).

\bibitem{Sato}
H. Takami, H. Yoshiguchi, K. Sato,  Astrophys.J. {\bf 639}, 803 (2006).

\bibitem{danny1}
D. De Marco, P. Blasi and A.O. Olinto, JCAP {\bf 01}, 002 (2006).

\bibitem{danny2}
D. De Marco, P. Blasi and A.V. Olinto, JCAP {\bf 07}, 015 (2006).

\bibitem{TT}
P. G. Tinyakov and I. I. Tkachev, JETP Lett. {\bf 74}, 445 (2001).

\bibitem{Auger}
P. Sommers \etal [The Pierre Auger Collaboration],
Proc. of 29th Int. Cosmic Ray Conf., {\bf 7}, 387 (2005),
astro-ph/0507150.

\bibitem{DBO}
D. De Marco, P. Blasi, A. Olinto, Astropart. Phys. {\bf 20}, 53 (2003).

\bibitem{DBO1}
D. De Marco, P. Blasi, A. Olinto, JCAP {\bf 01}, 002 (2005),
astro-ph/0507324.

\bibitem{Berez05}
V.~Berezinsky, astro-ph/0509069.

\bibitem{Teshima06}
M.~Teshima (for AGASA collaboration) talk at CRIS 2006 conference.

\bibitem{Blanton}
M.~Blanton, P.~Blasi, A.~Olinto, Astropart. Phys. {\bf 15}, 275 (2001).

\bibitem{Ueda}
Y.~Ueda, M.~Akiyama, K.~Ohta, T.~Miyaji, Astrophys. J.,{\bf 589},886 (2003).

\bibitem{Barger}
A.~J.~Barger \etal, Astron. J. {\bf 129}, 578 (2005).

\bibitem{Morris}
S.~L.~Morris \etal, Astrophys. J, {\bf 380}, 49 (1991).

\bibitem{blasirev}
P. Blasi, Mod. Phys. Lett. {\bf A20}, 3055 (2005)

\bibitem{BGH}
V.~Berezinsky, S.~Grigorieva, B.~Hnatyk, Astropart. Phys. {\bf 21}, 617 (2004).

\bibitem{michael}
M. Kachelriess and D. Semikoz, Phys. Lett. B {\bf 634}, 143 (2006).

\bibitem{Allard}
D. Allard et al, Astron. Astroph. {\bf 443}, L29 (2005).

\bibitem{berez}
E.G. Berezhko and L.T. Ksenofontov, J. of Exp. and Theor. Phys. {\bf
  89} 391 (1999)

\bibitem{Sigl}
G. Sigl and E. Armengaud, JCAP, {\bf 10}, 016 (2005), astro-ph/0507656 .

\bibitem{kascade}
K.-H. Kampert et al (KASCADE-Collaboration),  Proceedings of 27th ICRC,
volume "Invited, Rapporteur, and Highlight papers of ICRC", 240 (2001).

\bibitem{horandel}
J.R. H\"{o}erandel, {\it Preprint} astro-ph/0508014

\bibitem{parizot}
D. Allard, E. Parizot, A.V. Olinto, {\it Preprint} astro-ph/0512345

\bibitem{Emax}
E. G. Berezhko, Proc. of 27th ICRC (Hamburg), Invited, Rapporteur and
Higlight papers, 226 (2001),\\
V. S. Ptuskin and V. N. Zirakashvilli, Astron.Astroph. {\bf 429}, 755 (2005).

\bibitem{gabici}
S. Gabici and P. Blasi, Astroph. J. {\bf 583} 695 (2003)

\bibitem{blasi2001}
P. Blasi, Astropart. Phys. {\bf 15} 223 (2001)

\bibitem{malkov1}
M.A. Malkov, {\it Astrophys. J.} {\bf 485}, 638 (1997).

\bibitem{malkov2}
M.A. Malkov, P.H. Diamond P.H. and H.J. V\"{o}lk, {\it
Astrophys. J. Lett.} {\bf 533}, 171 (2000).

\bibitem{blasi1}
P. Blasi, {\it Astropart. Phys.} {\bf 16}, 429 (2002).

\bibitem{blasi2}
P. Blasi, {\it Astropart. Phys.} {\bf 21}, 45 (2004).

\bibitem{amato}
E. Amato and P. Blasi, MNRAS Lett. {\bf 364} 76 (2005);
E. Amato and P. Blasi, MNRAS {\bf 371} 1251 (2006).

\bibitem{bell78}
A.R. Bell, {\it MNRAS} {\bf 182}, 443 (1978).

\bibitem{vietri}
M. Vietri, {\it Astrophys. J.} {\bf 591}, 954 (2003).

\bibitem{blasivietri}
P. Blasi and M. Vietri, {\it Astrophys. J.} {\bf 626}, 877 (2005).

\bibitem{gallant}
A. Achterberg, Y. A. Gallant, J. G. Kirk and A. W. Guthmann,
MNRAS 328, 393 (2001).

\bibitem{lemsh}
M. Lemoine and B. Revenu, MNRAS {\bf 366}, 635 (2006).

\bibitem{morlino1}
G. Morlino, P. Blasi and M. Vietri, {\it Particle acceleration at
  shock waves moving at arbitrary speed: the case of large scale
  magnetic field and anisotropic scattering}, {\it Submitted to
  Astrop. J.}

\bibitem{morlino2}
G. Morlino, P. Blasi and M. Vietri, {\it Particle acceleration at
  shock waves: particle spectrum as a function of the equation of
  state of the shocked plasma}, {\it Submitted to Astrop. J.}

\bibitem{epstein}
P. Blasi, R.I. Epstein, and A.V. Olinto, Astroph. J. Lett. {\bf 533}
123 (2000)

\bibitem{arons}
J. Arons, Astroph. J. {\bf 589} 871 (2003)

\bibitem{pinch}
V. P. Vlasov, S. K. Zhdanov, B. A. Trubnikov, Soviet Physics of Plasma
{\bf 16}, 1457 (1990).

\bibitem{vannoni}
P. Blasi, S. Gabici and G. Vannoni, MNRAS {\bf 361} 907 (2005)

\bibitem{ellison}
D. Ellison, L. O'C. Drury and J-P. Meyer, Astroph. J. {\bf 487} 197 (1997);\\
J-P. Meyer, L. O'C. Drury and D. Ellison, Astroph. J. {\bf 487} 182 (1997)


\end{thebibliography}
\end{document}